\documentclass[a4paper,11pt]{article}
\usepackage{jheppub} 
\usepackage{xcolor}
\usepackage{amsmath}

\usepackage{tikz}
\usetikzlibrary{shapes.geometric}
\usepackage{graphicx}
\usepackage{subcaption}
\usepackage{float}
\usepackage{cleveref}

\usepackage{mathtools}
\DeclarePairedDelimiter\bra{\langle}{\rvert}
\DeclarePairedDelimiter\ket{\lvert}{\rangle}
\DeclarePairedDelimiterX\braket[2]{\langle}{\rangle}{#1\,\delimsize\vert\,\mathopen{}#2}

\def\Wtt{\widetilde{\mathcal{W}_{\tilde{t}}}}
\def\volWt{\mathrm{Vol}(\widetilde{\mathcal{W}_{\tilde{t}}})}
\def\volwt{\mathrm{Vol}({{w}_{\tilde{t}}})}

\definecolor{fgreen}{rgb}{0.13, 0.55, 0.13}

\usepackage{cleveref}
\preprint{ICTS-USTC/PCFT-25-42}
\arxivnumber{2510.05264} 

\title{Entanglement Entropy and Complexity in Dyonic Quantum Black Holes}

\author[a]{Sanhita Parihar}
\author[b,c]{and Gurmeet Singh Punia}

\affiliation[a]{Department of Physics, Indian Institute of Technology Hyderabad, \\ Kandi, Sangareddy, Telangana 502285, India}
\affiliation[b]{Interdisciplinary Center for Theoretical Study, \\
University of Science and Technology of China, Hefei, Anhui 230026, China}
\affiliation[c]{Peng Huanwu Center for Fundamental Theory, Hefei, Anhui 230026, China }

\emailAdd{sanhita.hepth@gmail.com, gurmeet96@ustc.edu.cn}

\abstract{In this work, we study the holographic entanglement entropy (HEE) and holographic complexity (HC) for three-dimensional dyonic quantum black holes, incorporating corrections arising from bulk quantum fields in the setup of double holography. We investigate the holographic entanglement entropy through the holographic Ryu-Takayanagi (RT) prescription and the island prescription. Using RT extremization, we evaluate HEE for connected and disconnected (island) surfaces and show islands emerge when RT surfaces intersect the brane; entanglement entropy grows with subregion size and ultimately saturates for quantum black holes as well as dressed defects. For complexity, we analyze both CV (perturbative) and CA (exact, all-orders) prescriptions: the leading quantum corrections feature universal behavior and the late-time growth can be expressed in thermodynamic variables, obeying generalized Lloyd-type bounds. In contrast, quantum dressed defects exhibit vanishing late-time growth. The CA prescription proves to be more tractable nonperturbatively and yields a thermodynamic interpretation of complexity growth.}

\begin{document}
\maketitle
\flushbottom

\section{Introduction}
The AdS/CFT conjecture offers a non-perturbative framework for quantum gravity in asymptotically Anti-de Sitter (AdS) spacetime by connecting it to a boundary conformal field theory (CFT) \cite{Maldacena:1997re, Witten:1998qj}. It has become a central tool in advancing our understanding of quantum gravity, although a complete picture of holography is still lacking. A significant avenue of exploration lies in understanding how spacetime emerges from the underlying degrees of freedom of the field theory. In this context, the Ryu-Takayanagi (RT) proposal has played a particularly pivotal role in advancing our understanding \cite{Ryu:2006ef, Hubeny:2007xt, PhysRevD.34.373, Srednicki_1993}. The proposal states that the entanglement entropy (measure of quantum information) of a subregion $\mathbf{R}$ in the $d$-dimensional CFT is encoded in the area of a codimension-two minimal surface in the bulk, which is homologous to the boundary region $\mathbf{R}$. 
\begin{align}
    S_{EE}(\mathbf{R)} = \frac{\text{min}({\cal{A}_{\mathbf{R}}})}{4G_N^{(d+1)}},
\end{align}
where \({\cal{A}_{\mathbf{R}}}\) is the area of the extremal surface anchored on the boundary of the subregion $\mathbf{R}$, and $G_{N}^{(d+1)}$ is the $(d+1)$ dimensional Newton's constant.  For time-dependent scenarios the proposal is generalized by Hubeny-Rangamani-Takayanagi (HRT) \cite{Hubeny:2007xt}, which considers:
\begin{align}
    S_{EE} = \frac{\text{min}(\text{extremal}~{\cal{A}})}{4G_N^{(d+1)}}.
\end{align}
While entanglement entropy captures essential features of spatial correlations, it is not suitable for describing the dynamics of the Einstein-Rosen (ER) bridge inside a black hole \cite{Susskind:2014moa}. 
In holography, it is conjectured that the growth of the ER bridge in an AdS black hole corresponds to the increasing complexity of the boundary quantum state's circuit \cite{Susskind:2014rva, Stanford:2014jda}. The computational complexity of a quantum state, defined as the number of unitary gates needed to reach the target state from an initial reference state, increases monotonically over time. This concept has led to the study of holographic complexity, considered dual to the quantum computational complexity of the boundary CFT state \cite{Susskind:2018pmk}.

The holographic dual of circuit complexity in quantum computation theory has been shown to be linked to gravitational observables. These include the regularized volume of the largest codimension-one surface crossing the Einstein-Rosen bridges (``{complexity = volume}''(CV) proposal) \cite{Susskind:2014rva, Stanford:2014jda}, the low-energy on-shell action of the Wheeler-DeWitt (WDW) patch (``complexity = action'' (CA) proposal)  \cite{Brown:2015bva, Brown:2015lvg}, and the volume of the WDW patch (`` complexity = spacetime volume'' (CV2.0) proposal) \cite{Couch:2016exn}. Recently, it has been found that there is an infinite class of diffeomorphism-invariant gravitational observables that display universal features of complexity, which has led to the ``complexity = anything'' proposal \cite{Belin:2021bga, Belin:2022xmt, Jorstad:2023kmq}. These proposals of holographic complexity have been studied in various holographic setups in AdS \cite{Carmi:2016wjl, Carmi:2017jqz, Chapman:2016hwi, Alishahiha:2015rta, Swingle:2017zcd, Goto:2018iay, An:2018dbz, Jiang:2019qea, Liu:2019mxz, Jorstad:2023kmq, Wang:2023ipy, Zhang:2024mxb, Emami:2024mes, Omidi:2022whq, Mandal:2022ztj, Jiang:2025qai, Jiang:2023jti, Wang:2023eep} as well as (asymptotically) de Sitter (dS) spacetime  \cite{Reynolds:2017lwq, Chapman:2021eyy, Aguilar-Gutierrez:2023pnn, Aguilar-Gutierrez:2024rka, Anegawa:2023wrk, Paul:2025gpk,Susskind:2021esx,Jorstad:2022mls,Baiguera:2023tpt,Anegawa:2023dad,Aguilar-Gutierrez:2023zqm}.

In this paper, we will follow the two proposals (CA \& CV) for holographic complexity. The CV conjecture proposes that complexity is dual to the volume of a maximal codimension-1 slice anchored on the boundary time slice $\Sigma_{CFT}$ on which the CFT state is defined,
\begin{equation}\label{eq:cv1.0def}
\mathcal{C}_V (\Sigma_{CFT}) = \underset{\Sigma_{CFT} = \partial \mathcal{B}}{\text{max}}  \left[ \frac{V(\mathcal{B})}{G_N \ell_{\text{bulk}} } \right] \,,
\end{equation}
where, $G_N$ denotes Newton's constant in the bulk gravitational theory, $\ell_{\text{bulk}}$ is a length scale associated with the bulk geometry and $V(\mathcal{B})$ is the volume of the codimension-one bulk surface $\mathcal{B}$ anchored to the boundary $\Sigma$. 

The CA conjecture proposes that the complexity of boundary states is proportional to the gravitational action evaluated in the Wheeler-De Witt (WDW) patch, \emph{i.e.} the region of union of all possible spacelike codimension-one surfaces anchored at constant boundary time. The CA proposal is then given by
\begin{equation}\label{eq:cadef}
\mathcal{C}_A = \frac{I_{\text{WDW}}}{\pi \hbar} \,.
\end{equation}
In all the discussion above, it was assumed that the bulk gravitational theory is purely classical \emph{i.e.} the central charge of the boundary CFT has been taken to be infinite. In this paper, we will take into account the quantum effects of bulk quantum fields on the holographic entanglement entropy\footnote{A similar study has been done in \cite{Chakrabortty:2020ptb, Pant:2025zhw} where the authors have consider the quantum backreaction by heavy-flavor insertions and boundary deformations on the HEE, HC and choas.} as well as holographic complexity in the Karch-Randall (KR) braneworld setup \cite{Karch:2000ct}, also known as \textit{double holography}. Double holography\footnote{A top-down string theory realization of double holography has been given in \cite{Uhlemann_2021, Karch_2022, Deddo:2023oxn}.} facilitates us with three different descriptions of a theory. 
\begin{itemize}
    \item \textbf{Boundary Perspective:} Describes a $d$-dimensional $\text{CFT}_{d}$ with a co-dimension one conformal defect\footnote{The case when the $\text{CFT}_{d}$ has a boundary on which the $(d-1)$-dimensional CFT lives give rise to AdS/BCFT \cite{Takayanagi:2011zk}.}.
    \item \textbf{Brane perspective:} Gives the holographic dual description of the boundary perspective ($\text{CFT}_{d}$ with defect) in terms of a matter-coupled $d$-dimensional dynamical gravity theory, which is joined to an external bath $\text{CFT}_{d}$ via transparent boundary conditions.
\item \textbf{Bulk perspective:} The $d$ dimensional matter coupled dynamical gravity with the bath $\text{CFT}_{d}$ has a $\text{AdS}_{d+1}$ gravity dual with a $\text{AdS}_{d}$ brane (End of the world brane) of tension $\tau$ on which the dynamical gravity is localized. The $\text{CFT}_{d}$ bath is the boundary dual theory sitting on the fixed boundary of $\text{AdS}_{d+1}$.
\end{itemize}
 The double holography setup transforms the semiclassical problem of solving the field equations with quantum back-reactions to a classical problem by switching from the brane perspective to the bulk perspective. In other words, the classical black holes in $\text{AdS}_{d+1}$ with the brane describe quantum corrected black holes in $\text{AdS}_{d}$ with all orders of quantum corrections \cite{Emparan:2002px, Emparan_2000, Emparan:1999fd}.  

In the context of entanglement entropy, the quantum corrections to the RT proposal have been found in \cite{Engelhardt_2015, Faulkner:2013ana}
\begin{eqnarray}
    S_{\mathbf{R}}=\text{ext}\left(\frac{A(\mathbf{X})}{4G_{N}}+S_{\text{ent}}\right),
\end{eqnarray}
where $A(\mathbf{X})$ is the area of the co-dimension two surface which is homologous to the boundary region $\mathbf{R}$ and $S_{\text{ent}}$ is the bulk entanglement entropy across the bulk extremal surface $\mathbf{X}$. It has been shown that this result can be understood using double holography \cite{Almheiri:2019hni}, and the entanglement entropy is reproduced by considering the RT surface in the higher-dimensional bulk theory. In this setup of double holography, there are two possible RT surfaces connected (Hartman-Maldacena surface) and disconnected (island surfaces). The disconnected RT surface corresponds to the extremal surface that intersects with the brane and gives rise to the islands (a region on the brane inside the horizon). This led to the island proposal\footnote{In KR setup due to the presence of bath, the AdS graviton become massive, for details on this refer to \cite{Geng:2020qvw}.} \cite{Penington:2019npb, Almheiri:2019psf}, the entropy of some region $\mathbf{R}$ of a holographic quantum field theory that is coupled to a gravity theory is given by,
\begin{eqnarray}
    S_{\text{gen}}=\text{Min}\left(\text{Ext}_{I}\left(\frac{A(\partial I)}{4G_{N}}+S(R\cup I)\right)\right),
\end{eqnarray}
where $A(\partial I)$ is the area of the boundary of the island.

While in \cite{Almheiri:2019hni} they have considered 2-dimensional gravity+matter theory, the results have been generalized to eternal black holes in five-dimensional Schwarzschild AdS black holes \cite{Almheiri:2019psy}, eternal charged Reissner-Nordstr\"om (RN) black holes in general dimensions \cite{Ling:2020laa}, and dyonic black holes \cite{Jeong:2023lkc}. Moreover, in \cite{Lin:2024gip}, the entanglement entropy of a 4-dimensional Reissner-Nordstr\"om black hole has been studied, and the page curve is reproduced using the island formula. Also, a study of entanglement entropy where the bulk spacetime is pure $\text{AdS}_{d+1}$, and the boundary CFT lives on $\mathbb{R}\times S^{d-1}$ has been presented in \cite{Chen:2020uac}.

In a similar spirit, in the setup of braneworld holography, the quantum corrections to holographic complexity\footnote{Also in \cite{Hernandez:2020nem} the authors have studied subregion complexity in a double holographic setup.} are explored for neutral black holes \cite{Emparan:2021hyr} as well as rotating black holes \cite{Chen:2023tpi}, following both CV and CA. The quantum corrections in the volume complexity can be distinguished into three different origins: (1) semiclassical backreaction on the classical black hole geometry $V(\Sigma)$, (2) higher derivative corrections to the gravity $v(\sigma)$, and (3) quantum corrections due to bulk quantum fields $C_{V}^{\text{bulk}}(\ket{\psi})$, while such a distinction cannot be achieved in the case of action complexity. Still, the action conjecture has been much more efficient in the braneworld setup, as it provides holographic complexity incorporating all order corrections due to bulk quantum fields. 
\begin{eqnarray}\label{eq: CV_quant}
    C_{V}&=&\frac{V(\Sigma)}{G_{N}^{(d+1)}\ell_{\text{bulk}}}+\frac{\delta V(\Sigma)+v(\sigma)}{G_{N}^{(d+1)}\ell_{\text{bulk}}}+C_{V}^{\text{bulk}}(\ket{\psi}), \nonumber\\
    C_{A}&=&\frac{I_{WDW}}{\pi \hbar}+\delta C_{A}(\ket{\psi}).
\end{eqnarray}


In this work, we build upon previous investigations of holographic complexity in neutral and rotating black holes \cite{Emparan:2021hyr, Chen:2023tpi}. We extend the analysis to dyonic charged quantum black hole geometries, introducing an unexplored probe of holographic entanglement entropy within the setup that has not been previously investigated. Another motivation to consider this system is that it serves as a practical model for exploring magneto-transport phenomena and quantum Hall states in strongly coupled quantum theories that operate under different chemical potentials and curved background geometries. These aspects have wide applications in condensed matter physics. More broadly, this framework allows for the study of defects that can carry both electric charge and magnetic flux, which could potentially model anyonic excitations.

Before proceeding with the computation, we summarize the central findings of this work. The quantum black hole geometries inherently encode universal quantum corrections arising from boundary matter backreaction, and we examine their implications for both HEE and HC. In the case of HEE, the study reveals the emergence of two types of extremal RT surfaces - connected and disconnected (island) surfaces within the double holographic framework. For both quantum dyonic black holes $(\kappa=-1)$\footnote{The notation will be explained clearly in the following section.} as well as quantum dressed conical defects $(\kappa = 1)$, the entanglement entropy increases monotonically with subregion size and eventually saturates. The appearance of disconnected RT surfaces intersecting the brane exhibits the formation of islands, and the minimization of HEE with island size supports the island proposal. In addition, we have studied the holographic complexity of quantum dyonic black holes and quantum-dressed charged conical defects, employing both the volume and action conjectures of complexity. Within this setup, the volume prescription restricts us to a perturbative treatment of the leading quantum corrections, whereas the action prescription captures exact non-perturbative contributions. Finally, we interpret the resulting complexity growth in terms of thermodynamic quantities, drawing parallels with the saturated Lloyd bound for classical dyonic black holes, while extending the analysis to a broader set of thermodynamic variables. Moreover, for quantum dressed defects, we get a vanishing complexity growth, consistent with extremal Reissner–Nordström behavior.

This paper is organized as follows: section \ref{sec:qbh} presents a review of three-dimensional quantum dyonic black holes in a braneworld setup. In section \ref{sec:EE}, we study the holographic entanglement entropy for both connected and disconnected phases in this setup, with all orders of quantum backreaction. In section \ref{sec:CV} and \ref{sec:AC}, we study the holographic complexity of three-dimensional quantum dyonic black holes following CV and CA proposals, respectively. In conclusion, we summarize our findings and outline the strategic direction for our future initiatives in section \ref{sec:discussion}.
 
\section{ Quantum charged black holes}\label{sec:qbh}
In this section, we present the main features of the quantum charged black hole. A three-dimensional quantum charged black hole can be constructed through the KR braneworld approach by placing an $\text{AdS}_3$ brane with tension $\tau$ within the four-dimensional bulk spacetime described by a four-dimensional AdS charged C-metric. Following the notation of \cite{Climent:2024nuj, Feng:2024uia}, the total action of the system is expressed as:
\begin{equation}
    I = I_{\text{bulk}}[\mathcal{M}] + I_{\text{GHY}}[\partial \mathcal{M}] + I_{\text{brane}}[\mathcal{B}] \,,
\end{equation}
where the bulk action, Gibbons-Hawking-York (GHY) action, and the action of the brane are respectively given by
\begin{align}\label{eq:actionbulk}
\begin{aligned}
    & I_{\text{bulk}} = \frac{1}{16\pi G_4} \int_{\mathcal{M}} d^4x \sqrt{-g} \left( \mathcal{R} + \frac{6}{\ell_4^2} \right) - \frac{1}{4g_{\star}^2} \int_{\mathcal{M}} d^4x \sqrt{-g} F_{\mu\nu}F^{\mu\nu}\,, \\[2pt]
    & I_{\text{GHY}} = \frac{1}{8\pi G_4} \int_{\partial \mathcal{M}} d^3x \sqrt{-h} K\,, \\[2pt]
    & I_{\text{brane}} = -\tau \int_{\mathcal{B}} d^3x \sqrt{-h}\,,
\end{aligned} 
\end{align}
where $\mathcal{R}$ denotes the Ricci scalar of the bulk spacetime $\mathcal{M}$ equipped with the metric $g_{\mu\nu}$, $K$ is the extrinsic curvature scalar defined on the asymptotic boundary $\partial \mathcal{M}$, and $h$ represents the induced metric on the brane $\mathcal{B}$. The constant $G_4$ is the four-dimensional Newton's constant. The electromagnetic field tensor $F$ couples minimally to the spacetime curvature via a dimensionless gauge coupling constant $g_{\star}$, with $\ell_{\star}^2 \equiv 16\pi G_4 / g_{\star}^2$. Solutions within this framework are holographically dual to states of a $(2+1)$-dimensional conformal field theory (CFT) characterized by chemical potentials and a central charge given by $c_{3} = \ell_4^2 / G_4$. All the details of the solution are provided in the following section.

\subsection{Review of the charged AdS C-metric \& thermal quantities}
In this section, we outline the key characteristics of the charged extension of the quantum BTZ (quBTZ) black hole \cite{Climent:2024nuj, Feng:2024uia}. The $\mathrm{AdS}_4$ C-metric, which includes both electric and magnetic charges of the bulk black hole, originates from a four-dimensional solution of the Einstein-Maxwell-AdS theory. This four-dimensional setup describes a three-dimensional quantum black hole on a brane placed in the bulk within the context of braneworld holography. The metric of the four-dimensional charged AdS black hole can be conveniently written in the following form:
\begin{align}\label{eq:met-sol}
\mathrm{d} s^2 & =\frac{\ell^2}{(\ell + x r)^2} \left[-H(r) \mathrm{d} t^2+\frac{ \mathrm{d} r^2}{H(r)}+r^2\left(\frac{ \mathrm{d} x^2}{G(x)}+G(x) \mathrm{d}\phi^2\right)\right]\,,\\[4pt] 
H(r)&=\frac{r^2}{\ell_3^2}+\kappa-\frac{\mu \ell}{r}+\frac{q^2 \ell^2}{r^2}\,, \qquad G(x)=1-\kappa x^2-\mu x^3 -q^2 x^4\,, 
\end{align}
 and the gauge field takes the following form\footnote{The normalization is fixed such that the electromagnetic field is minimally coupled with gravity with a coupling constant $g_{\star}$ and it has been regularized at the horizon $r_{+}$ and $x_{1}$}:
\begin{align}\label{eq:gauge-sol}
    \mathbf{{F}} & =\mathrm{d}{\mathcal{A}}\,,\,\quad {\mathcal{A}} = -\frac{2\ell}{\ell_{\star}}\left[e \left( \frac{1}{r_+}-\frac{1}{r}\right)\mathrm{d} t + g(x-x_1) \mathrm{d} \phi\right]\,,
\end{align}
where $e$ and $ g$ are related to the electric and magnetic charges of the black hole \footnote{The bulk charges are given by $$ Q_e = \frac{2}{g_\star^2} \int \star F = \frac{8 \pi \ell e \Delta x_1}{g_\star^2 \ell_\star} \,, \qquad Q_g = \frac{2}{g_\star^2} \int F = \frac{8 \pi \ell g \Delta x_1}{g_\star^2 \ell_\star},$$ and $q^2 = e^2 + g^2 $.}, and $\kappa$ is a dimensionless discrete parameter that takes values $\kappa = \pm 1,0$ corresponding to three dimensional quantum-dressed conical defects, quantum black holes, and Poincare $\text{AdS}_{3}$ on the brane, respectively. Note that we have chosen a gauge potential such that it is regular at the relevant zeroes of $H(r)$ and $G(x)$, namely $H(r_+) = 0$ and $G(x_1) = 0$, where $r_+$ is the largest positive root and $x_1$ is the smallest positive root of the respective polynomial. To ensure a Lorentzian signature, we need to impose $G(x) \geq 0$. Therefore, we will consider only the portion of the spacetime where $0 \leq x \leq x_1$. The regularity at $G(x_{1})=0$ constrain the continuous real parameter $\mu$ to satisfy,
\begin{equation}
    \mu=\frac{1+x_{1}^{2}-q^{2}x_{1}^{2}}{x_{1}^{3}}.
\end{equation}
This metric satisfies the Einstein equation
\begin{equation}\label{rejop3988}
R_{\mu\nu}+2 {F}_\mu{}^{\rho} {F}_{\rho \nu}+\frac{1}{2} {F}_{\rho\sigma}  {F}^{\rho\sigma}  g_{a b}=-3\left(\frac{1}{\ell^2}+\frac{1}{\ell_3^2}\right) g_{\mu\nu}\equiv -\frac{3}{\ell_4^2}g_{\mu\nu}\,,
\end{equation}
where 
\begin{equation}\label{eq:l4def}
\frac{1}{\ell_4^2}=\left(\frac{1}{\ell^2}+\frac{1}{\ell_3^2}\right) \,.
\end{equation}
We place a brane at $x=0$ in the bulk spacetime \eqref{eq:met-sol}, and truncate the bulk at the position of the brane. Then, to complete the spacetime, we glue an identical copy of the bulk at the brane position. The Israel junction condition across the brane $K^{+}_{\mu\nu}-K^{-}_{\mu\nu}=\frac{\partial S}{\partial h_{\mu\nu}}$ fix the brane tension \cite{Israel:1966rt}
\begin{equation}
    \tau = \frac{1}{2 \pi G_4 \ell},
\end{equation}
which provides the bulk interpretation of $\ell$. Thus, the induced three-dimensional metric on the brane is given by,
\begin{equation}
\mathrm{d} s^2=-H(r) \mathrm{d} t^2+\frac{\mathrm{d} r^2}{H(r)}+r^2 \mathrm{d} \phi^2\,,
\end{equation}
This metric can be seen as a solution to the three dimensional effective theory of gravity coupled to the gauge field on the brane \cite{Climent:2024nuj}, and therefore it is interpreted as the quantum backreacted black holes, often referred as ``\emph{quantum black holes}''.

Then, we have a final regularity condition considered for the smoothness of the geometry along the rotational symmetry axis $x=x_1$, and the absence of conical singularities requires that the coordinate $\phi$ be periodically identified as
\begin{align}
    \phi \sim \phi+2 \pi \Delta \quad \text { with } \quad \Delta=\frac{2}{\left|G^{\prime}\left(x_1\right)\right|}=\frac{2 x_1}{\left|-3+\kappa x_1^2-q^2 x_1^4\right|}
\end{align}
After a canonical renormalization of coordinates by rescaling
\begin{align}
    t=\Delta \bar{t}, \quad \phi=\Delta \bar{\phi}, \quad r=\frac{\bar{r}}{\Delta} \,.
\end{align}
The metric on the brane is given by,
\begin{align}
    \left.d s^2\right|_{x=0}=-H(\bar{r}) d \bar{t}^2+\frac{d \bar{r}^2}{H(\bar{r})}+r^2 d \bar{\phi}^2 \,,
\end{align}
with
\begin{align}
    H(\bar{r})=\frac{\bar{r}^2}{\ell_3^2}-8 \mathcal{G}_3 M-\frac{\ell F(M, q)}{\bar{r}}+\frac{\ell^2 Z(M, q)}{\bar{r}^2} \,,
\end{align}
and
\begin{align}
    \begin{aligned}
    & F(M, q)=\mu \Delta^3=8 \frac{1-\kappa x_1^2-q^2 x_1^4}{\left(3-\kappa x_1^2+q^2 x_1^4\right)^3} \,, \\
    & Z(M, q)=q^2 \Delta^4=q^2 \frac{16 x_1^4}{\left(-3+\kappa x_1^2-q^2 x_1^4\right)^4} \,,
    \end{aligned}
\end{align}
where $\mathcal{G}_3$ is `renormalized' three-dimensional Newton’s constant, given by
\begin{align}
    \mathcal{G}_3 = \frac{\ell_4}{\ell} G_3 = \frac{1}{2\ell} G_4 \,,
\end{align}
and $M$ is the mass of the black hole. Following \cite{Kudoh:2004ub, Emparan_2006, Climent:2024nuj, Feng:2024uia}, one can express the mass in terms of the conical deficit as\footnote{Notice that for the case of black holes \emph{i.e.} $\kappa=-1$, $M>0$ and for the quantum dressed conical defects \emph{i.e.} $\kappa=1$ the mass is negative $M<0$.} 
\begin{align}
    M = - \frac{\kappa}{8 G_3} \frac{\ell}{\ell_4} \Delta^2 \,.
\end{align}
The temperature and generalized entropy associated with the three-dimensional black hole are
\begin{align}
    T &= \frac{\Delta H'(r_+)}{4\pi} 
    = \frac{3\Delta}{4\pi x_1 \ell_3} 
    \left[ \frac{1}{z} + \frac{\kappa z x_1^2}{3} - \frac{\nu^2 q^2 x_1^4 z^3}{3} \right], \\
    S_{\text{gen}} &= \frac{2}{4 G_4} \int_0^{2\pi \Delta} d\phi \int_0^{x_1} 
    \frac{\ell^2 r_+^2}{(\ell + x r_+)^2} dx 
    = \frac{\pi \ell_3 \Delta \sqrt{1 + \nu^2}}{2 G_3 x_1 z (1 + \nu z)},
\end{align}
where we have defined
\begin{equation}
    z = \frac{\ell_3}{r_{+} x_1}\,, \quad \gamma = q x_1^2, \quad \gamma_e = e x_1^2, \quad \gamma_g = g x_1^2.
\end{equation}
With these identifications, the other thermodynamic parameters then take the form as
\begin{align}
    M &= \frac{\sqrt{1 + \nu^2} \left( z \nu + 1 \right) \left( 1 + \gamma^2 \nu^2 z^4 + \nu \left( \gamma^2 - 1 \right) z^3 \right) z^2}
    {2 G_3 \left( 1 + \gamma^2 \nu^2 z^4 + (2 \gamma^2 + 2) \nu z^3 + (\gamma^2 + 3) z^2 \right)^2}, \\
    T &= \frac{(2 - \gamma^2 \nu^3 z^5 - 2 \gamma^2 \nu^2 z^4 - \nu (\gamma^2 + 1) z^3 - 3 z \nu) z}
    {2 \pi \ell_3 \left( 1 + \gamma^2 \nu^2 z^4 + (2 \gamma^2 + 2) \nu z^3 + (\gamma^2 + 3) z^2 \right)}, \\
    S_{\text{gen}} &= \frac{\pi \ell_3 \sqrt{1 + \nu^2} z}
    {G_3 \left( 1 + \gamma^2 \nu^2 z^4 + (2 \gamma^2 \nu + 2 \nu) z^3 + (\gamma^2 + 3) z^2 \right)}, \\
    \mu_{e,g} &=\sqrt{ \frac{5 g_3^2}{4 \pi G_3} } \frac{\nu \gamma_{e,g} z^3 (z \nu + 1)}
    {(1 + \gamma^2 \nu^2 z^4 + (2 \gamma^2 + 2) \nu z^3 + (\gamma^2 + 3) z^2)}, \\
    Q_{e,g} &= \sqrt{ \frac{16\pi}{5 g_3^2 G_3}}  \frac{\gamma_{e,g} z^2 (z\nu + 1) \sqrt{1 + \nu^2}}
        {(1 + \gamma^2 \nu^2 z^4 + 2\nu (\gamma^2 + 1) z^3 + (\gamma^2 + 3) z^2)}.
\end{align}
 Taken all together, these quantities satisfy the first law of thermodynamics
 \begin{align}
     dM = T dS_{\text{gen}} + \mu _e dQ_e + \mu_g dQ_g,
 \end{align}
where the variations are computed at fixed values of $G_3, g_3 , \ell_3$ and $\nu$.

\noindent In addition to the standard thermodynamic quantities, it has been shown in \cite{Kubiznak:2012wp, Kubiznak:2016qmn, Frassino:2022zaz} that one can also introduce a notion of thermodynamic pressure, defined as
\begin{equation}
P_{4}=-\frac{\Lambda_{4}}{8\pi G_{4}}=\frac{3(\nu^{2}+1)}{8\pi G_{4}\ell^{2}}.
\end{equation}
Correspondingly, the associated thermodynamic volume takes the form
\begin{equation}
V_{4}=\frac{8 \pi  \ell^3 z^2 (2 \nu  z+1)}{3 \nu ^2 \left(\nu ^2 \gamma ^2 z^4+2 \nu  \left(\gamma ^2+1\right) z^3+\left(\gamma ^2+3\right) z^2+1\right)^2}.
\end{equation}
Moreover, if the brane tension is treated as a thermodynamic variable by allowing variations in the backreaction parameter $\ell$, one obtains a corresponding thermodynamic area of the brane given by,
\begin{eqnarray}
A_{\tau}=\frac{2 \pi  \ell^2 z^2 (\nu  z+1) \left(\nu  \left(\nu +z \left(\nu  z \left(\nu  \gamma ^2 z (\nu  z+1)+\nu  z+2\right)-2\right)\right)-2\right)}{\left(\nu +\nu  z^2 \left((\gamma +\nu  \gamma  z)^2+2 \nu  z+3\right)\right)^2}.
\end{eqnarray}
Here we keep $G_{4}$ fixed while allowing $\ell$ and $\ell_{3}$ to vary. With these extensions, the first law of thermodynamics generalizes to
\begin{equation}
dM=TdS_{\text{gen}}+\mu_{e} dQ_{e}+\mu_{g} dQ_{g}+V_{4}dP_{4}+A_{\tau}d\tau.
\end{equation}

\section{Entanglement Entropy}\label{sec:EE}

In this section, we study the entanglement entropy in our setup of double holography, where the holographic conformal field theories with a chemical potential is coupled to the 3-dimensional charged quantum black hole which is joined to a $\text{CFT}_{3}$ bath with transparent boundary conditions, and the bulk dual is charged $\text{AdS}_{4}$ C-metric with a brane.
\begin{figure}[h!]
\begin{minipage}{0.45\textwidth}
    \centering
    \begin{tikzpicture}[scale=1.5]
    \draw[thick] (0,0) circle (1.5);
    \draw[black] (-2.2,0) -- (1.6,0);
    \draw[very thick,blue] (-0.02,1.51) arc(48.5:-48.5:2cm);
    
    \draw[rotate=90,very thick,red] (-1,1.11) parabola bend (0,0) (1,1.11);
    
    \fill[red!50, opacity=0.5] plot[domain=0:1.1,samples=100] (-\x,{0.95*sqrt(\x)}) -- (-1.1,-1.02) arc(42.5:-42.5:-1.5cm)  -- plot[domain=1.1:0,samples=100] (-\x,{-0.95*sqrt(\x)}) -- cycle;
    
    \begin{scope}[shift={(0.4,0)}] 
    \draw[very thick] plot[domain=0:0.17,samples=100] (\x,{1.4*sqrt(\x)});
    \draw[very thick] plot[domain=0:0.17,samples=100] (\x,{-1.4*sqrt(\x)});
    \fill[blue!50, opacity=0.6] 
    plot[domain=0:0.17,samples=100] (\x,{1.4*sqrt(\x)})  
    -- (0.22,0.36) 
    arc(10.5:-10.5:2cm)                                 
    -- plot[domain=0.17:0,samples=100] (\x,{-1.4*sqrt(\x)}) 
    -- cycle;
    \end{scope}
    
    \filldraw[red] (-1.5,0) circle (1pt);
    \filldraw[red] (0,0) circle (1pt);

    \node at (0.8,0.3) {$r_0$};
    \node at (0.8,-0.95) {\small $x=0$};
    \node at (0.5,1.1) {\small $\mathcal{B}$};
    \node at (-2.1,0.1) {$\phi$};
    \node at (-1.6,0.2) {$x_1$};
    \node at (0.2,0.15) {\small $(x_\star,r_\star)$};
    \node at (-1.4,0.9) {$\epsilon$};
    \draw [->] (-2,0.2) to [out=60,in=30] (-2,-0.2);
    \end{tikzpicture}
\end{minipage}
\hfill
\begin{minipage}{0.45\textwidth}
    \centering
    \begin{tikzpicture}[scale=1.5]
    \draw[thick] (0,0) circle (1.5);
    \draw[black] (-2.2,0) -- (1.6,0);
    \draw[very thick,blue] (-0.02,1.51) arc(48.5:-48.5:2cm);
    
    
    \begin{scope}[shift={(0.64,0)}] 
    \draw[very thick,red] plot[domain=0:1.4,samples=100] (-\x,{0.3+0.5*(\x)*(\x)});
    \draw[very thick,red] plot[domain=0:1.4,samples=100] (-\x,{-0.3-0.5*(\x)*(\x)});
    
    \fill[red!50, opacity=0.5] 
    plot[domain=0:1.4,samples=100] (-\x,{0.3+0.5*(\x)*(\x)}) 
    -- (-1.37, 1.3) 
    arc[start angle=-61,end angle=61,radius=-1.5cm] 
    -- plot[domain=1.4:0,samples=100] (-\x,{-0.3-0.5*(\x)*(\x)}) 
    -- cycle;
    \end{scope}
    
    \begin{scope}[shift={(0.41,0)}] 
    \draw[very thick] plot[domain=0:0.17,samples=100] (\x,{1.4*sqrt(\x)});
    \draw[very thick] plot[domain=0:0.17,samples=100] (\x,{-1.4*sqrt(\x)});
    \fill[blue!50, opacity=0.6] 
    plot[domain=0:0.17,samples=100] (\x,{1.4*sqrt(\x)})  
    -- (0.22,0.36) 
    arc(10.5:-10.5:2cm)                                 
    -- plot[domain=0.17:0,samples=100] (\x,{-1.4*sqrt(\x)}) 
    -- cycle;
    \end{scope}
    
    \filldraw[red] (-1.5,0) circle (1pt);
    \filldraw[blue] (0.63,0.3) circle (1pt);
    \filldraw[blue] (0.63,-0.3) circle (1pt);
    \filldraw[red] (0,0.51) circle (1pt);

    \node at (0.8,-0.95) {\small $x=0$};
    \node at (0.5,1.1) {\small $\mathcal{B}$};
    \node at (0.8,0.3) {$r_0$};
    \node at (-2.1,0.1) {$\phi$};
    \node at (-1.6,0.2) {$x_1$};
    \node at (0.0,0.8) {\small $(x_c,r_c)$};
    \node at (-1.4,0.9) {$\epsilon$};
    \draw [->] (-2,0.2) to [out=60,in=30] (-2,-0.2);
    \end{tikzpicture}
\end{minipage}
    \caption{This schematic diagram shows the RT surface in the Karch-Randall brane $\mathcal{B}$ at $x=0$ (thick blue line), enclosing a three-dimensional black hole (shaded blue region) with a thick black line representing the Poincaré disk of the $\text{AdS}_{4}$-C metric.  \emph{Left:} Connecting RT surface, \emph{Right:} Disconnected surface intersecting brane at fixed $r_0$ (in shaded blue region) represent the region inside the black hole.}
\label{fig:connanddisc}
\end{figure}
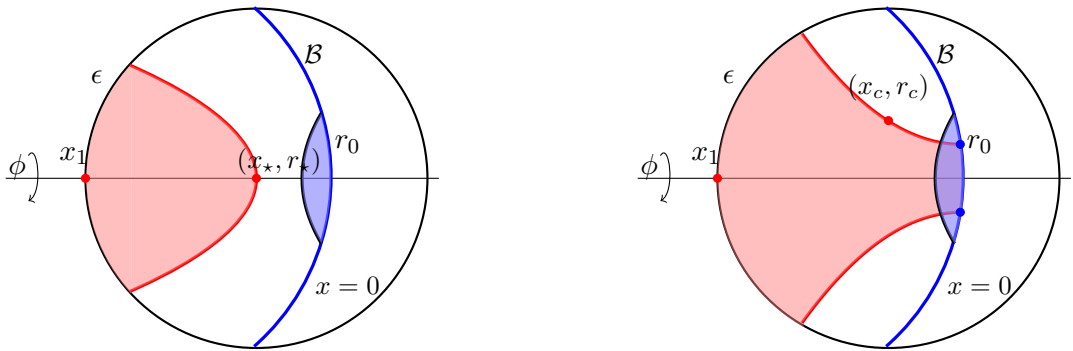
The boundary of spacetime described by $\text{AdS}_{4}$ C-metric is at $r=r_{b}$ with boundary coordinates $\{t,x,\phi\}$\footnote{The asymptotic boundary metric is given by,
\begin{equation}
    ds^{2}=-H(x) dt^2+\ell^2  \left(\frac{1}{x^{2}G(x)}+\frac{1}{x^{4}H(x)}\right)dx^2+\frac{\ell^2 G(x) }{x^2}d\phi^2,
\end{equation}
There is a singularity at $x=0$ hidden under the horizon described by $H(x)=0$.}, with $0<x<x_{1}$ and $0<\phi<2\pi\Delta$. Let us consider a boundary region of radius $\epsilon$, around $x_{1}$, and find the entanglement entropy of this boundary region by following the RT/island prescription for both connected and disconnected RT surfaces as shown in figure \ref{fig:connanddisc}.

To proceed with the calculation of entanglement entropy, first we find the pullback metric on a co-dimension two surface with embedding vector $X^{\mu}=\{t(x),r(x),x,\phi\}$\footnote{Notice that a general embedding should be $X^{\mu}=\{t(r(x),x),r(x),x,\phi\}$. We did not work with this for computational simplification, as the extremization of the area functional becomes quite challenging even numerically. Therefore, our results are not the most general, but we expect them to be valid for the static spacetime, which is the case here. },
\begin{equation}
    ds^{2}=\frac{\ell^2}{(\ell+r x)^2} \left( \frac{ \left(H(r) \left(r^2-G(x) H(r) t'(x)^2\right)+G(x) r'(x)^2\right)}{G(x) H(r)}dx^2+r^2 G(x) d\phi^2\right),
\end{equation}
then the area functional is given by,
\begin{eqnarray}
    A&=&2\int d\phi\int^{x_{1}}_{\epsilon} dx \sqrt{h} \,, \nonumber\\
    &=&4\pi \Delta \int^{x_{1}}_{\epsilon} dx \frac{\ell^2 r \sqrt{H(r) \left(r^2-G(x) H(r) t'(x)^2\right)+G(x) r'(x)^2}}{\sqrt{H(r)} (\ell+x r(x))^2} .\nonumber
\end{eqnarray}
 As the area functional does not depend on $t(x)$ due to the static nature of spacetime, the associated momentum is conserved,
\begin{equation}
    \frac{\partial A}{\partial t'(x)}=\text{constant}.
\end{equation}
 This conserved momentum has been utilized to find that $t'(x)=0$ extremizes the area functional. So, the area functional simplifies,
 \begin{eqnarray}\label{eq:EE_connected}
    A&=&4\pi \Delta \int_{\epsilon}^{x_{1}} dx\frac{\ell^2 r \sqrt{r'(x)^2 G(x)+r(x)^2 H(r)}}{\sqrt{H(r)} (\ell+x r(x))^2}.
\end{eqnarray}
Now, if we think of this area functional as a Lagrangian and coordinate $x$ as time, then this is equivalent to a system that is explicitly dependent on time $\mathcal{L}=\mathcal{L}(r(x),x,r'(x))$. Therefore, there is no conserved Hamiltonian to help out with the extremization of the Lagrangian/ area functional. So, we directly proceed to the Euler-Lagrange equation. Unfortunately, the Euler-Lagrange equation is a second-order non-linear partial differential equation, and we cannot solve it analytically. But we can proceed numerically by solving Euler's equation with fixed boundary conditions.
\begin{figure}[H]
\begin{minipage}{0.45\textwidth}
    \centering
    \includegraphics[width=\textwidth]{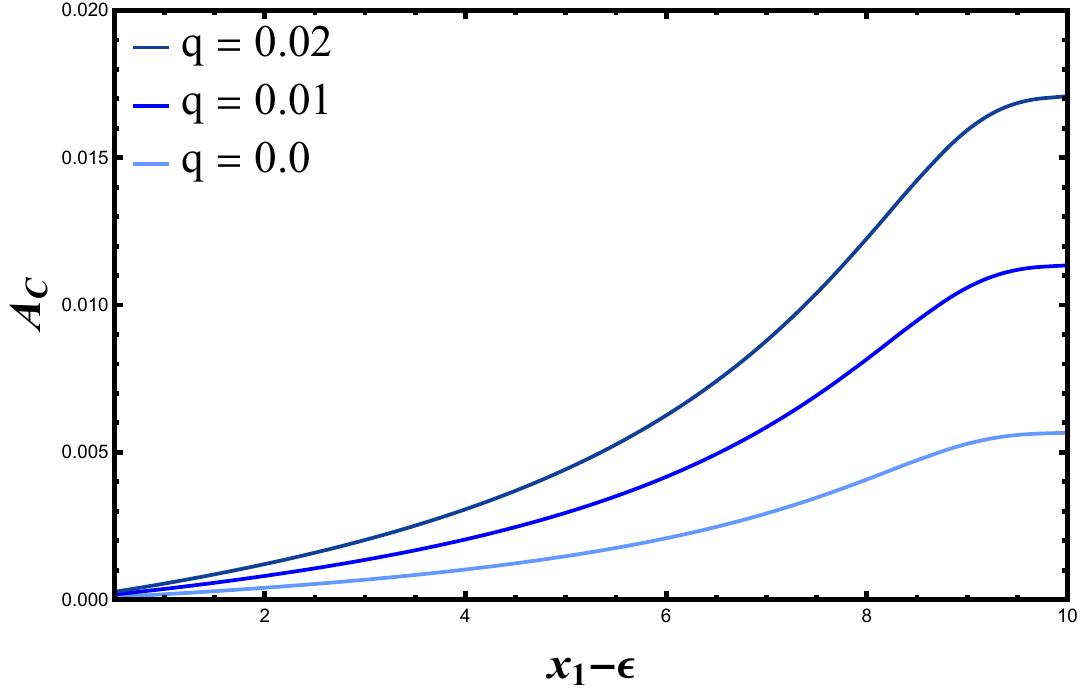}
\end{minipage}\label{fig:Avsepsilon1}
\hfill
\begin{minipage}{0.45\textwidth}
    \centering
    \includegraphics[width=\textwidth]{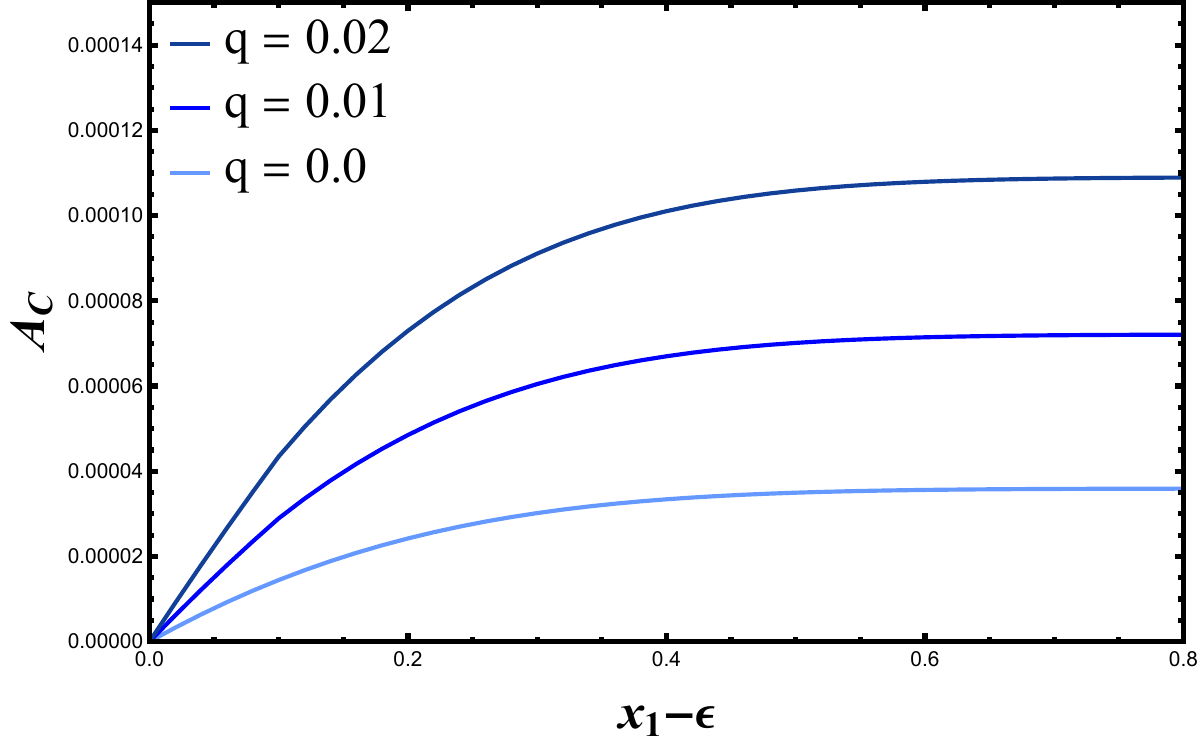}
\end{minipage}
    \caption{\emph{Left:} $A_{C}$ vs subregion radius $(x_{1}-\epsilon)$ for $\kappa=-1$, \emph{Right:} $A_{C}$ vs subregion radius $(x_{1}-\epsilon)$ for $\kappa=1$.}
\label{fig:Avsepsilon-1}
\end{figure}
 For the connected RT surface we fix the boundary conditions at the extremum point, $r'(x)=0$ at $x=x_{\star}$ and $r(x_{\star})=r_{\star}$, and find the area functional numerically in terms of $x_{\star}$ and $r_{\star}$. 
 \begin{eqnarray}\label{eq:EE_connected2}
    A_{C}(x_{\star},r_{\star})&=&4\pi \Delta \int^{x_{1}}_{\epsilon} dx\frac{\ell^2 r \sqrt{r'(x)^2 G(x)+r(x)^2 H(r)}}{\sqrt{H(r)} (\ell+x r(x))^2}.
\end{eqnarray}
We illustrate the numerical behavior of entanglement entropy\footnote{While the exact expression of entanglement entropy is $S_{E}=\frac{A}{4G_{4}}$, we have plotted an $\frac{A}{4\pi\Delta}=\frac{G_{4}S_{E}}{\pi \Delta}$ for the sake of simplicity.} as a function of the boundary subregion length $(x_{1}-\epsilon)$ in figure \ref{fig:Avsepsilon-1} for both $\kappa=\pm1$ (at fixed values of $(x_{\star},r_{\star})$ such that entanglement entropy is minimized). As it can be seen from figure \ref{fig:Avsepsilon-1} for $\kappa=\pm1$, corresponding to quantum black holes and quantum dressed defects in $\text{AdS}_{3}$ respectively, the area increases monotonically with subregion length but eventually saturates. The key difference we observe is that the HEE saturates faster in quantum dressed defect compared to the quantum black hole. 

\begin{figure}[h!]
\begin{minipage}{0.45\textwidth}
    \centering
    \includegraphics[width=\textwidth]{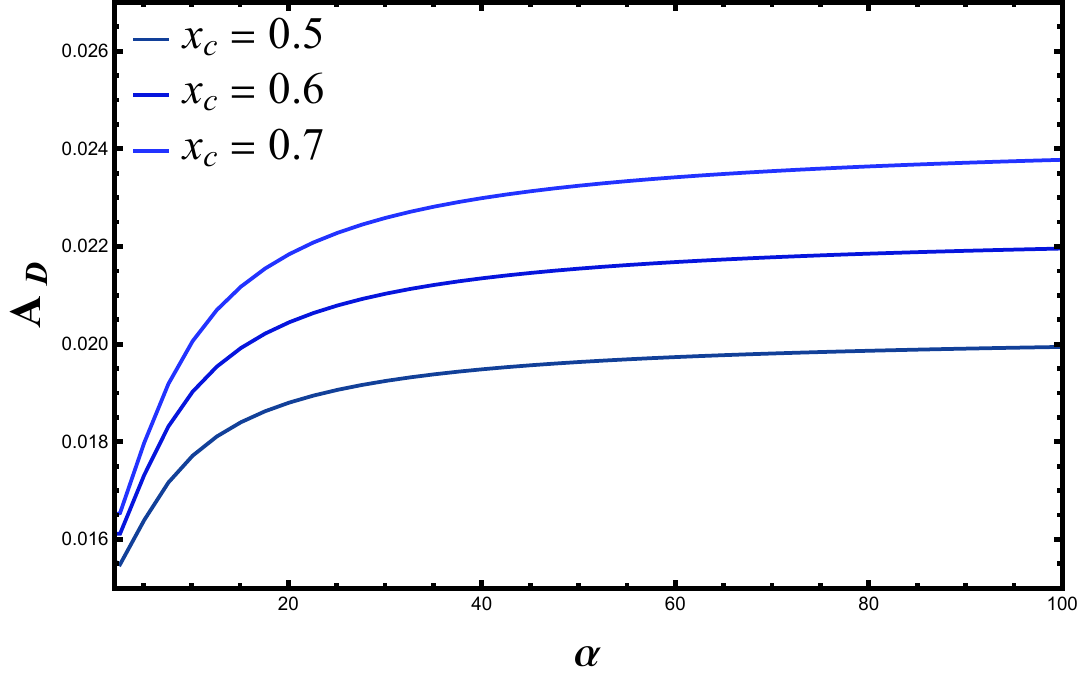}
\end{minipage}
\hfill
\begin{minipage}{0.45\textwidth}
    \centering
    \includegraphics[width=\textwidth]{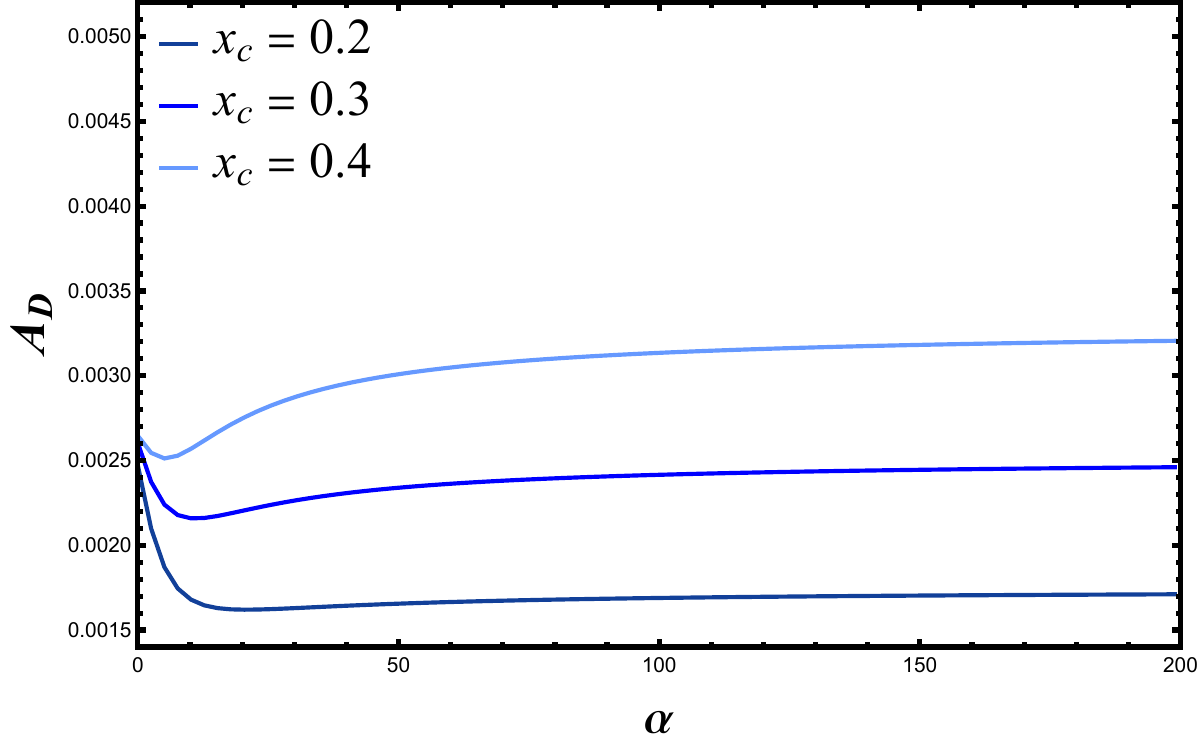}
\end{minipage}
    \caption{\textit{Left:} $A_{D}$ vs $\alpha$ for $\kappa = -1$, $q=0$, \textit{Right:} $A_{D}$ vs $\alpha$ for $\kappa =1$, $q=0$ and different value of $x_c$'s}
    \label{fig:Avsalpha}
\end{figure} 
To study the disconnected RT surface, we again consider the same boundary region of radius $\epsilon$ around $x_{1}$ as we did for the connected RT surface. While for the disconnected RT suface we have divided the RT surface between two regions from $\{0,x_{c}\}$ and $\{r_{c},r_{b}\}$, with $x(r_{c})=x_{c}$ and vice versa, following \cite{Almheiri:2019psy}. This uplifts the ambiguity if the RT surface is multi-valued, and the area functional is given by,
\begin{eqnarray}
    A_{D}(r_{c},x_{c})&=&\int d\phi\int_{r_{c}}^{r_{b}} dr \frac{\ell^2 r \sqrt{G(x(r),q)+r^2 x'(r)^2 H(r,q)}}{(\ell+r x(r))^2 \sqrt{H(r,q)}},\nonumber\\&&+\int d\phi\int^{x_{c}}_{0} dx \frac{\ell^2 r(x) \sqrt{r'(x)^2 G(x,q)+r(x)^2 H(r(x),q)}}{(\ell+x r(x))^2 \sqrt{H(r(x),q)}},\nonumber\\ 
\end{eqnarray}
Let us say, $r_{0}$ is the point where the RT surface intersects the brane, for a schematic diagram, refer to figure \ref{fig:connanddisc}. This intersecting point at the brane $r_{0}$ can be determined by the solution of the  Euler-Lagrange equation in terms of $r(x)$, by evaluating it at the brane position $r(0)=r_{0}$. 
\begin{figure}[h!]
\begin{minipage}{0.31\textwidth}
    \centering
    \includegraphics[width=\textwidth]{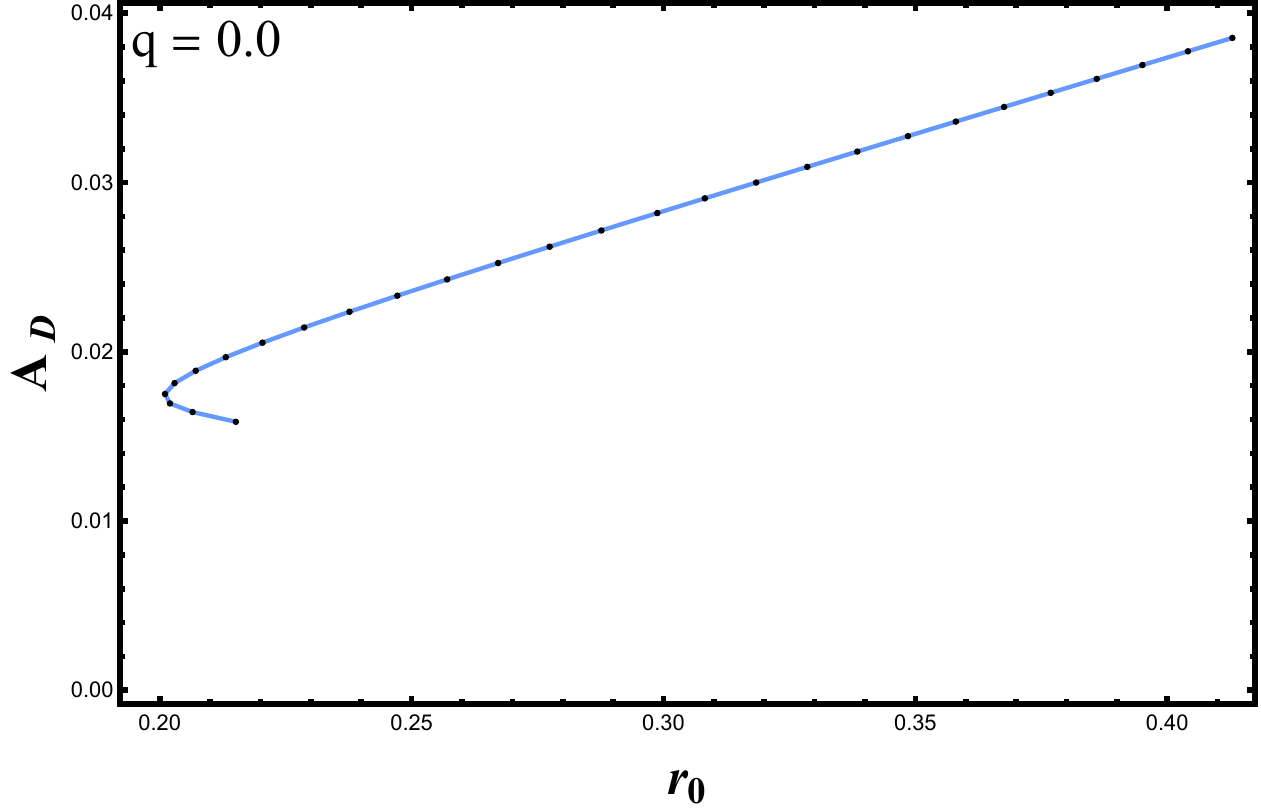}
    \subcaption{}
    \label{fig:4a}
\end{minipage}
\hfill
\begin{minipage}{0.31\textwidth}
    \centering
    \includegraphics[width=\textwidth]{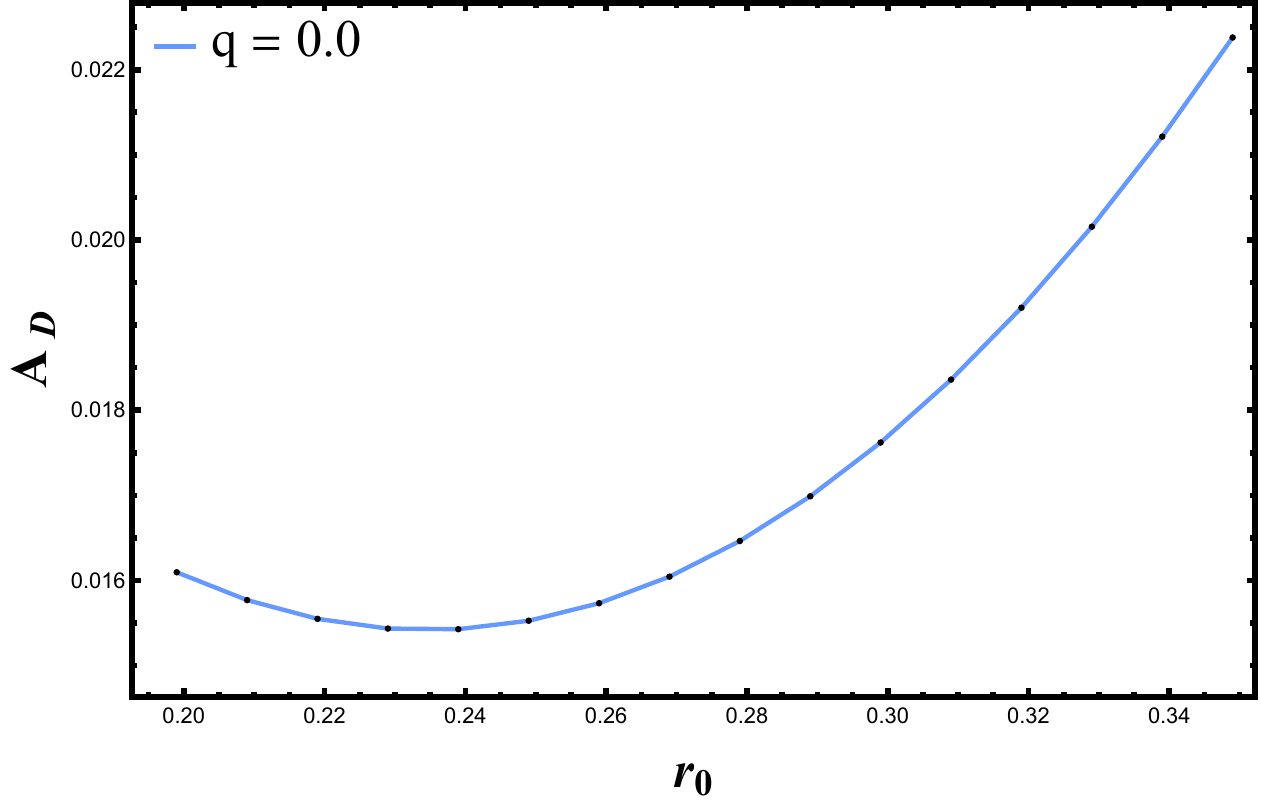}
    \subcaption{}
    \label{fig:4b}
\end{minipage}
\hfill
\begin{minipage}{0.31\textwidth}
    \centering
    \includegraphics[width=\textwidth]{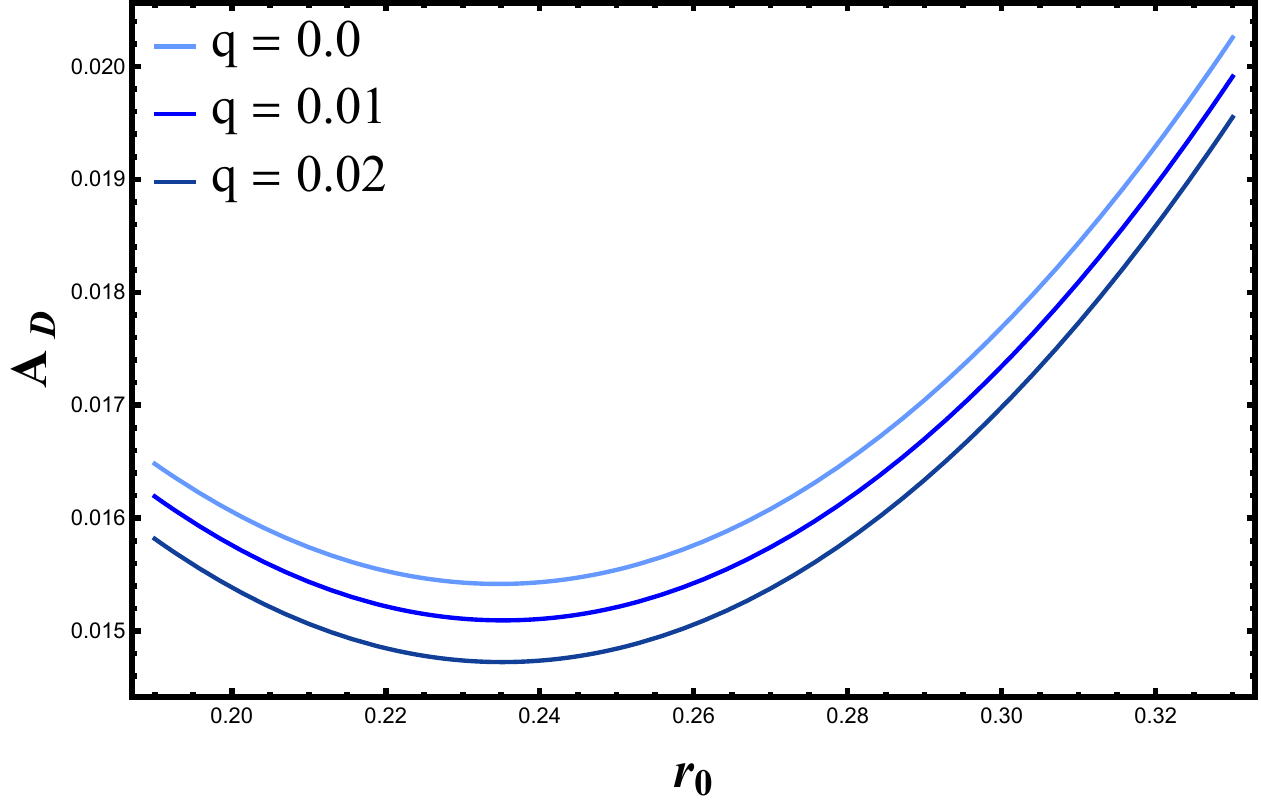}
    \subcaption{}
    \label{fig:4c}
\end{minipage}
\caption{\textbf{(a)} The area profile of the disconnected surface $A_D$ is shown as a function of the intersection point $r_0$, with $x_c$ varied and $r_c$ and $\alpha$ minimized, for $\kappa = -1$ and $q = 0$. \textbf{(b)} The area profile of $A_D$ is plotted as a function of $r_0$, varying $\alpha$ while minimizing $x_c$ and $r_c$, exhibits a shallow dip. \textbf{(c)} The plots in (b) are now shown for several non-zero values of charges.}
    \label{fig:Avsr0k-1}
\end{figure}
We have done the extremization for both parts of the disconnected area functional separately with the boundary conditions, $x(r_{c})=x_{c}$ and $x'(r_{c})=\alpha$ for the first part and $r(x_{c})=r_{c}$ and $r'(x_{c})=(x'(r_{c}))^{-1}$ for the second part. After the extremization procedure, we have the area of the disconnected RT surface as a function of $x_{c}$, $r_{c}$, and $\alpha$. 

\begin{figure}[H]
\begin{minipage}{0.31\textwidth}
    \centering
    \includegraphics[width=\textwidth]{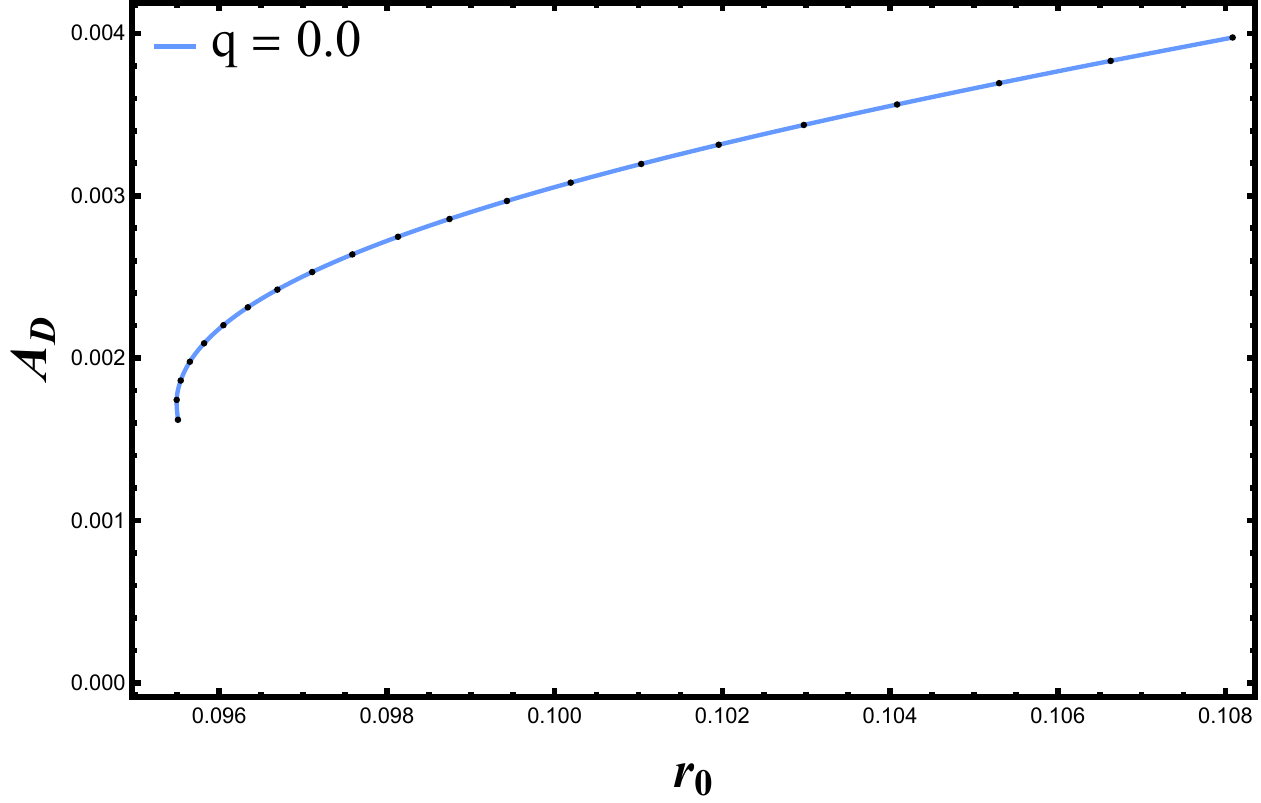}
    \subcaption{}
    \label{fig:5a}
\end{minipage}
\hfill
\begin{minipage}{0.31\textwidth}
    \centering
    \includegraphics[width=\textwidth]{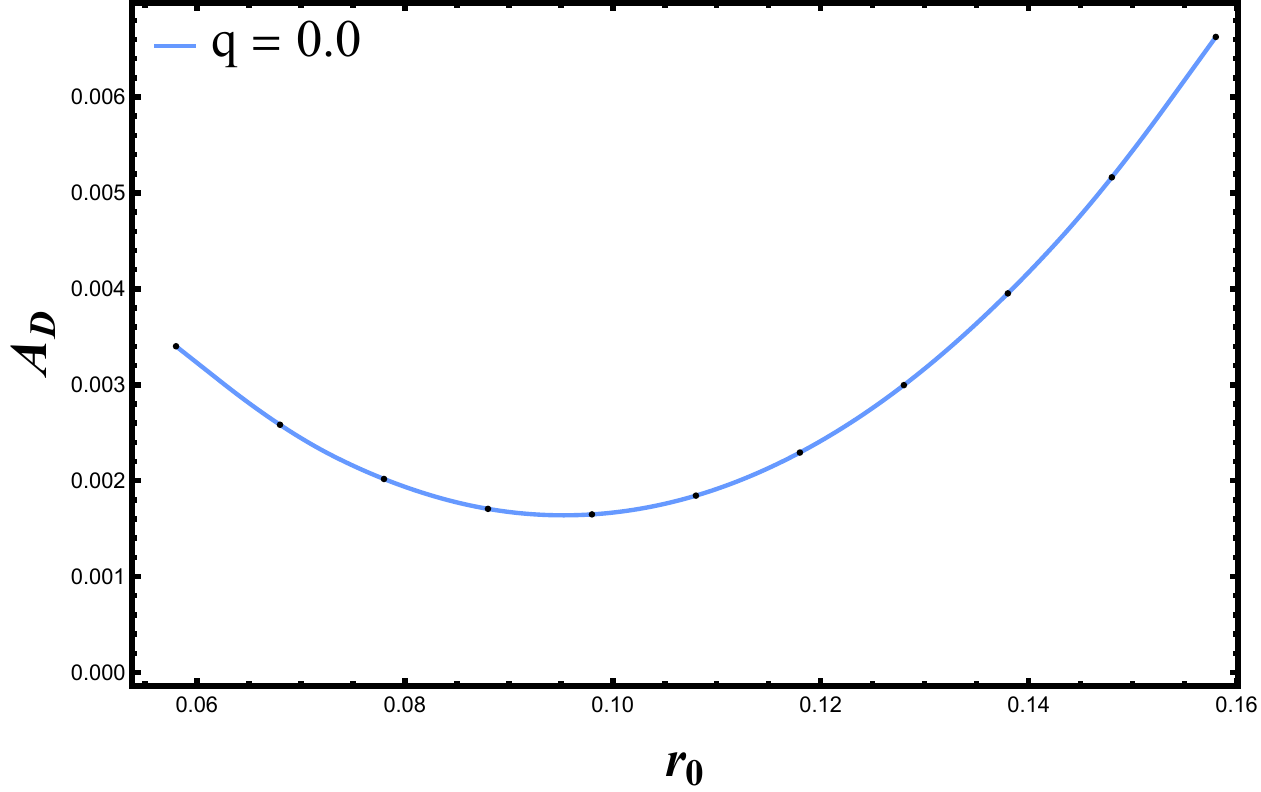}
    \subcaption[]{}
    \label{fig:5b}
\end{minipage}
\hfill
\begin{minipage}{0.31\textwidth}
    \centering
    \includegraphics[width=\textwidth]{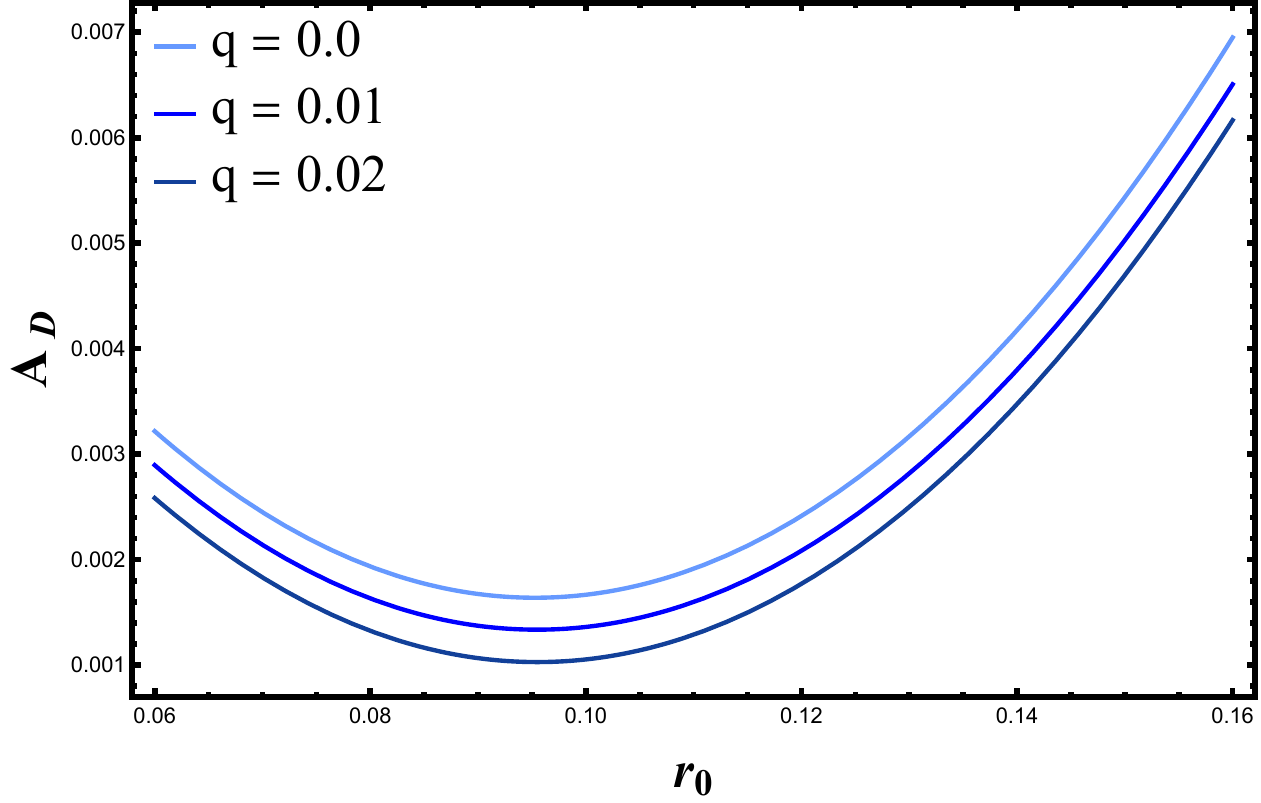}
    \subcaption{}
    \label{fig:5c}
\end{minipage}
    \caption{\textbf{(a)} The area profile of the disconnected surface $A_D$ is shown as a function of $r_0$, with $x_c$ varied and $r_c$ and $\alpha$ minimized, for $\kappa = 1$ and $q = 0$. \textbf{(b)} The area profile of $A_D$ is plotted as a function of $r_0$, varying $\alpha$ while minimizing $x_c$ and $r_c$. \textbf{(c)} The plots in (b) shown for several non-zero values of $q$.} 
    \label{fig:Avsr0k1}
\end{figure}

Among these three parameters, we first fix the parameter $\alpha$, so that the area is minimized. The explicit behavior of the area with $\alpha$ is shown in figure \ref{fig:Avsalpha}. This leaves us with $A_{D}$ as a function of two remaining parameters. Furthermore we observe that the area is a monotonic function of the parameters $x_{c}$ and $r_{c}$, this enables us to fix either one of them so that area is minimized and we can study the parametric dependence of area $A_{D}$ on the intersection point $r_{0}$ by varying the other parameter. In figure  \ref{fig:4a} and  \ref{fig:5a} we have shown a parametric plot of $A_{D}(r_{0})$ by varying $x_{c}$ and fixing $(\alpha, r_{c})$ at their minimizing values. For both signs of $\kappa$, the resulting curve is almost monotonic on the full range of $r_{0}$, and we do not see any explicit minima in these plots. But, if we choose to fix the parameters $(x_{c},r_{c})$ and study the parametric plot with respect to $\alpha$, this enables us to see a minima explicitly, they are presented\footnote{The plots shown in figure 4-5(a) and 4-5(b) are different parametric slices of $A_{D}(x_{c},r_{c},\alpha)$ and need not to match pointwise just they must agree near the extremum point given in terms of $r_{0}(x_{c},r_{c},\alpha)$.} in figure \ref{fig:4b} and  \ref{fig:5b}.
 At last, in figures \ref{fig:4c} and \ref{fig:5c}, we have shown the parametric behavior of $A_{D}$ with the intersection point $r_{0}$ for different values of charge parameters $q$, emphasizing how charge shifts the minima. As shown in figure \ref{fig:Avsr0k-1} and \ref{fig:Avsr0k1}, the area functional attains a minimum with respect to $r_{0}$ for both $\kappa=\pm1$. Therefore as $r_{0}$ determines the intersection point with the brane (or the island region), this depicts the validity of the island proposal for the quantum black holes.


\section{Volume complexity}\label{sec:CV}
Following the standard prescription, the volume complexity of the time-evolved thermo-field double state $\ket{\text{TFD}_{\beta}(t)}$ of the CFT is determined by the extremal volume of a spacelike slice $\Sigma_{t}$ in the bulk as given in \eqref{eq:cv1.0def}.  The volume complexity of the brane system with this standard prescription also incorporates the quantum back-reaction from the matter on the brane, so the \eqref{eq:cv1.0def} admits an expansion to study the quantum correction in the volume of the extremal surface as \eqref{eq: CV_quant}. This expansion can be understood in terms of the parameters of the theory, by introducing an effective coupling $g_{\text{eff}}$ as,
\begin{equation}
    g_{\text{eff}}\sim\frac{|I_{\text{CFT}}|}{|I_{\text{grav}}|}\sim\frac{G_{3}c_{3}}{\ell_{3}} \,,
\end{equation}
which in the limiting case of $\ell<<\ell_{3}$ simplifies to,
\begin{eqnarray}
    g_{\text{eff}}\sim\frac{\ell}{\ell_{3}}<<1.
\end{eqnarray}
So, the quantum corrections to the volume complexity can be understood as an expansion in either $g_{\text{eff}}$ or $\ell$
\begin{align}
    \mathcal{C}_V {(t)} = \mathcal{C}_V^{\text{BTZ}} {(t)} + \mathcal{C}_V^q {(t)} + \mathcal{O}(g_{\text{eff}}^{2}) \,.
\end{align}
where $\mathcal{C}_V^{\text{BTZ}}$ denotes the classical BTZ black hole which gives the leading contribution, and $\mathcal{C}_{V}^q$ captures the leading quantum correction. 
\subsection*{Extremal volume}
Let us write the volume functional in \eqref{eq:cv1.0def} for our specific solution \eqref{eq:met-sol}. It will be convenient to define the coordinate $ z=x r$, so that the brane lies at $z=0$ and the asymptotic boundary is at $z=-\ell$. With this choice, we consider the embedding vector as $X^\mu=(t, r, z, \phi)$. Using the axial symmetry, we can restrict the extremization problem to the family of axially-symmetric hypersurfaces, $\partial_\phi X^\mu=\delta_\phi^\mu$, and the extremal surface $\Sigma_t $ parameterized as $t = t(z,r)$. Thus we have $y^1=r$ and $y^2=z$ as the two remaining parameters, and define $\dot{X}^\mu=\partial_r X^\mu$ and $X^{\mu \prime}=\partial_z X^\mu$. With this parametrization, the induced metric on the spacelike hypersurface reads as
\begin{align}
\begin{aligned}
    d s_{\Sigma}^2=\frac{\ell^2}{(\ell+z)^2} & {\left[\left(-H \dot{t}^2 + H^{-1} + \frac{z^2}{G r^2}\right) d r^2 + \left(-H t^{\prime 2} + G^{-1} \right) d z^2 +\right.} \nonumber \\
    & \left.-2\left(H \dot{t} t^{\prime} + \frac{z}{G r}\right) d r d z + r^2 G d \phi^2\right] \,.
\end{aligned}
\end{align}
The volume complexity (VC) \eqref{eq:cv1.0def} of the system is then obtained by finding the extremal value of the volume functional\footnote{In general the limits of integral $\int dr dz=\int_{-\infty}^{-\frac{\ell}{x}}dr(\int_{-\infty}^{-\ell}dz+\int_{0}^{r x_{1}}dz)+\int_{0}^{\infty}dr(\int_{-\infty}^{-\ell}dz+\int_{0}^{r x_{1}}dz)$ the first part of integral does not contribute as we are anchoring the constant time slice at $r_{\infty}$ rather than the asymptotic boundary. This part corresponds to the complexity of $\text{CFT}_{3}$ on the asymptotic boundary of $\text{AdS}_{4}$ \emph{i.e.} rigid bath contribution, which is UV divergent. Also, as in $\ell\rightarrow0$ limit $z\in\{-\infty\}\cup\{0,\infty\}$ the integral $\int dr dz=\int_{0}^{\infty}dr(\int_{0}^{r x_{1}}dz)$. This part is arises due to the short-range correlation between $\text{CFT}_{3}$ in a rigid bath and the asymptotic region of brane, and in general, this contributes a constant in complexity due to the static nature of the spacetime.}
\begin{align}
    \mathcal{C}_V(t)=\operatorname{ext}\left\{\frac{\ell \Delta}{G_4 \ell_3} \int d r d z \frac{\ell^2 r}{(\ell+z)^3} \sqrt{-H\left(\dot{t}+\frac{z t^{\prime}}{r}\right)^2+H^{-1}-G t^{\prime 2}}\right\} \,.
\end{align}
where we have considered $\ell_{\text{bulk}}=4\pi \ell_{3}$, and a factor of two is added for the reflection symmetry across the brane. Now, the generic way to proceed is to extremize the above volume functional, which leads to a second-order non-linear partial differential equation (PDE). To proceed further, we need to solve this PDE either analytically or numerically, which turns out to be a challenging task in itself. Therefore, we only intend to proceed order by order in the backreaction parameter.

As the volume functional in the $\ell\rightarrow0$ limit can be recast as,
\begin{eqnarray}
    \mathcal{C}_{V}(t)=2\int_{r_{\text{min}}}^{\infty}dr\int_{0}^{r x_{1}}dz \frac{\ell^{2}}{(\ell+z)^{3}}f(r,z),
\end{eqnarray}
where a factor of 2 is added due to two asymptotic boundaries. By using integration by parts:
\begin{eqnarray}\label{eq: totalCV}
    \mathcal{C}_{V}(t)&=&2\int_{r_{\text{min}}}^{\infty}dr\left( -\frac{\ell^{2}}{2(\ell+z)^{2}}f(z)\Big|_{0}^{r x_{1}}-\frac{\ell^{2}}{2(\ell+z)}\partial_{z}f(z)\Big|_{0}^{r x_{1}}+\frac{1}{2}\int_{0}^{rx_{1}}dz \frac{\ell^{2}}{(\ell+z)}\partial^{2}_{z}f(z)\right),\nonumber\\
    &=&\int_{r_{\text{min}}}^{\infty}dr \Big( f(0)+\ell~\partial_{z}f(0)-\frac{\ell^{2}}{r^{2}x_{1}^{2}}f(rx_{1})-\frac{\ell^{2}}{r x_{1}}\partial_{z}f(rx_{1})\nonumber\\&&+\int_{0}^{rx_{1}}dz \frac{\ell^{2}}{(\ell+z)}\partial^{2}_{z}f(z)\Big).
\end{eqnarray}
\subsection*{Classical contribution:}
The first term gives the classical contribution,
\begin{eqnarray}
    \mathcal{C}_{V}^{(0)}=\operatorname{ext}\left(\frac{ \Delta}{2G_3 \ell_3}\int_{r_{\text{min}}}^{\infty}dr~r\sqrt{-H_{0}\dot{t}^{2}+H^{-1}_{0}}\right),
\end{eqnarray}
where we have used the boundary condition on the brane at $z=0$, \emph{i.e.} $t'(r,0)=0$ and $\frac{\ell \Delta}{G_4 \ell_3}=\frac{ \Delta}{2G_3 \ell_3}+\mathcal{O}(\ell^{2})$. Now, if we parameterize the extremal surface with parameter $\lambda$,
\begin{eqnarray}
    \mathcal{C}_{V}^{(0)}=\operatorname{ext}\left(\frac{ \Delta}{2G_3 \ell_3}\int_{\lambda_{\infty}^{L}}^{{\lambda_{\infty}^{R}}}d\lambda~r \sqrt{-H_{0}(r)\dot{t}^{2}+H^{-1}_{0}(r)\dot{r}^{2}}\right),
\end{eqnarray}
with Dirichlet boundary conditions, $r(\lambda^{L,R}_{\infty})=r_\infty$ and $t(\lambda^{L,R}_{\infty})=\tilde{t}$.

We can proceed with the extremization of this volume functional. For the extermination, we can think of the volume functional as a Lagrangian $\mathcal{L}(\dot{t},\dot{r},r)$ which is in general a function of $r(\lambda)$ and $t(\lambda)$ but as in this case it is independent of $t$ we will have a constant of motion,
\begin{eqnarray}
    P_{t}(r)=\frac{\partial\mathcal{L}}{\partial \dot{t}}=-\frac{r H_{0}(r)^{3/2} \dot{t}}{\sqrt{\dot{r}^2-H_{0}(r)^2 \dot{t}^2}}.
\end{eqnarray}
As there is an additional reparameterization symmetry, $\mathcal{L}(\lambda)=\mathcal{L}(\tilde{\lambda})$, we make a choice of $\lambda$ such that $\mathcal{L}=1$, this simplifies,
\begin{equation}
    P_{t}(r)=-r^{2}H_{0}(r)\dot{t},
\end{equation}
and $\mathcal{L}=1$ gives,
\begin{equation}
   \dot{r}= \frac{\sqrt{H_{0}(r)r^2+P_{t}^2}}{r^2}.
\end{equation}
These two equations can be combined,
\begin{eqnarray}
    dt&=&-\frac{P_{t}dr}{H_{0}(r) \sqrt{r^2 H_{0}(r)+P_{t}^2}} \,,\nonumber\\ 
    2\int_{0}^{\tilde{t}}dt&=&-2\int_{r_{\text{min}}}^{r_{\infty}}dr\frac{P_{t}}{H_{0}(r) \sqrt{r^2 H_{0}(r)+P_{t}^2}}\,,\nonumber\\
    \tilde{t}&=&-\int_{r_{\text{min}}}^{r_{\infty}}dr\frac{P_{t}}{H_{0}(r) \sqrt{r^2 H_{0}(r)+P_{t}^2}}\,,
\end{eqnarray}
where $r_{\text{min}}$ is the value of $r$ where the extremal surface $t(r)$ attains its extremum,
\begin{equation}
     \frac{dr}{d\lambda}\Bigg|_{r=r_{\text{min}}}=0,\quad P_{t}=r_{\text{min}}\sqrt{-H_{0}(r_{\text{min}})}.
\end{equation}
The complexity in leading order in $\ell$ is given by,
\begin{eqnarray}\label{eq:zerothCV}
    \mathcal{C}_{V}^{(0)}&=&\frac{\Delta}{2G_{3}\ell_{3}}\int_{\lambda_{\infty}^{L}}^{{\lambda_{\infty}^{R}}}d\lambda, \nonumber\\
    &=&\frac{\Delta}{G_{3}\ell_{3}}\int_{r_{\text{min}}}^{r_{\infty}}dr \frac{r^{2}}{\sqrt{H_{0}(r)r^2+P_{t}^2}}.
\end{eqnarray}
where, $H_{0}(r)=\frac{r^{2}}{l_{3}^{2}}+\kappa$. If we consider $\kappa=-1$, we get the analog of the classical BTZ black hole result \cite{Carmi:2017jqz} with the classical horizon at $r_{+}=\ell_{3}$ hidding the charged defect. Below, we focus on the quantum correction and late-time growth of complexity in this case. For the $\kappa=1$ case, one can proceed similarly, with the only difference coming from the fact that there is no horizon classically \emph{i.e.} there is no real root of $H(r)$. The horizon arises purely from quantum backreaction; we deal with the case separately. 

\subsection*{Leading quantum correction:}
In this section, we consider the leading order quantum corrections, which gives $\mathcal{O}(\ell)$ contributions in complexity \emph{i.e.} the second term in \eqref{eq: totalCV}, given by
\begin{eqnarray}
   C_{V}^{\text{bulk}}(\ket{\psi})&=& \int_{r_{\text{min}}}^{\infty}dr (\ell~\partial_{z}f(0))\nonumber\\
   &=&\frac{\Delta}{2 G_{3} \ell_{3} \sqrt{-H(r) \dot{t}(r,0)^2-t'(r,0)^2+\frac{1}{H(r)}}}  \Big(-t'(r,0) (r \dot{G}(0,r) t'(r,0)\nonumber\\&&+2 H(r) \dot{t}(r,0)+2 r t''(r,0))-2 r H(r) \dot{t}(r,0) \dot{t}'(r,0)\Big),
\end{eqnarray}
this term vanishes due the boundary condition on the brane $t'(r,0)=0$ and $G(0)=1$. Physically, it can be understood as the quantum corrections due to bulk matter fields \cite{Emparan:2021hyr}, and in this particular case, it vanishes due to the boundary condition on the brane.

There will be additional $\mathcal{O}(\ell)$ quantum corrections, from \eqref{eq:zerothCV} which can be interpreted as semi classical corrections to the bulk extremal surface due to matter fields $\delta V(\Sigma)$: one from the correction to $H(r)=H_{0}(r)-\frac{\mu \ell}{r}$, another from the correction in $r_{\text{min}}=r_{\text{min}}^{(0)}+r_{\text{min}}^{(1)}$
\begin{eqnarray}
    C_{V}^{(1)}&=&\frac{\Delta}{2G_{3}\ell_{3}}  \int_{r_{\text{min}}^{(0)}+r_{\text{min}}^{(1)}}^{r_{\infty}} dr~\left(\frac{r^2 (\ell \mu  r-2 P_t^{0}\delta P_t )}{2 \left(r^2 H_{0}(r)+({P_t}^0)^2\right)^{3/2}}+\frac{r^2}{\sqrt{r^2 H_{0}(r)+(P_t^{0})^2}}\right),\nonumber\\
    &=&\frac{\Delta}{2G_{3}\ell_{3}}  \int_{r_{\text{min}}^{(0)}}^{r_{\infty} }dr~\frac{r^2 (\ell \mu  r-2 P_t^{0}\delta P_{t} )}{2 \left(r^2 H_{0}(r)+(P_{t}^{0})^2\right)^{3/2}}-\frac{\Delta}{2G_{3}\ell_{3}}\frac{r^2}{\sqrt{r^2 H_{0}(r)+(P_{t}^{0})^2}}r_{\text{min}}^{(1)}.
\end{eqnarray}
For initial time $\tilde{t}=0$, $r_{\text{min}}^{(1)}=\frac{\mu \ell}{2}$ and 
\begin{equation}
 2 P_{t}^{0}\delta P_{t} =2r_{\text{min}}^{(0)}r_{\text{min}}^{(1)}-\frac{4(r_{\text{min}}^{(0)})^{3}r_{\text{min}}^{(1)}}{\ell_{3}}+\ell\mu r_{\text{min}}^{(0)}.   
\end{equation}
where, $\delta P_{t}$ is the $\mathcal{O}(\ell)$ correction to $P_{t}$. Using these explicit expressions for initial times, the complexity simplifies to
\begin{equation}
    C_{V}^{(1)}=-\frac{\mu\ell\Delta}{4G_{3}},\qquad \mu=\frac{1+x_{1}^{2}-q^{2}x_{1}^{2}}{x_{1}^{3}},\qquad \Delta=\frac{2 x_{1}}{3+x_{1}^{2}+q^{2}x_{1}^{4}}.
\end{equation}
We notice that the leading order quantum correction to the complexity is the same as that of a neutral black hole \cite{Emparan:2021hyr}, as it does not depends explicitly on $q$. This is justified as explicit charge $q$ only appears at $\mathcal{O}(\ell^{2})$ in \eqref{eq:met-sol}, we do not expect to see any explicit charge $q$ dependence at $\mathcal{O}(\ell)$. At this order, the only charge dependence can be seen through parameters $\mu$ and $\Delta$.
\subsection{Late time growth of complexity}
In this section we focus on the late time regime and study the complexity growth. At late times, $r_{\text{min}}$ is determined by extremizing the conserved momentum in $r$ \cite{Barbon:2019tuq, Barbon:2020olv, Barbon:2020uux},
\begin{equation}
    \frac{dP_{t}(r)}{dr}\Bigg|_{r=r_{\text{min}}}=0.
\end{equation}
We can proceed order by order again, and determine $r_{\text{min}}$ up to $\mathcal{O}(\ell)$,
\begin{equation}
    r_{\text{min}}=\frac{\ell_{3}}{\sqrt{2}}+\ell\frac{\mu}{4},
\end{equation}
which gives,
\begin{equation}
P_t=\frac{\ell_{3}}{2}+\ell\frac{\mu}{\sqrt{2}},
\end{equation}
The complexity growth rate is determined by using the Hamilton-Jacobi equation for $\mathcal{L}$,
\begin{eqnarray}\label{eq:comprate}
    \frac{d v(\sigma)}{dt}=2\pi\Delta\left(\frac{\partial\mathcal{L}}{\partial \dot{t}}\Bigg|_{r=r_{\infty}^{R}}+\frac{\partial\mathcal{L}}{\partial \dot{t}}\Bigg|_{r=r_{\infty}^{R}}\right)=4\pi\Delta \times P_{t}(r_{\text{min}}).
\end{eqnarray}
The late time growth of complexity at $\mathcal{O}(\ell^{0})$ is given by,
\begin{equation}
    \frac{dC_{V}^{(0)}}{d\tilde{t}}=\frac{4\pi \Delta}{8\pi G_{3}\ell_{3}} P_{t}=\frac{\Delta}{4G_{3}}.
\end{equation}
Using $\Delta^{2}=8G_{3}M$, the classical piece is given by,
\begin{equation}
    \frac{dC_{V}^{(0)}}{d\bar{t}}=\frac{\Delta^{2}}{4G_{3}}=2M.
\end{equation}
The leading order correction to the late-time growth of complexity,
\begin{equation}
    \frac{dC^{(1)}_{V}}{d\bar{t}}=\frac{\Delta ^2 \ell \mu }{2 \sqrt{2} G_{3} \ell_{3}}=2\sqrt{2}\frac{\mu\ell}{\ell_{3}}M.
\end{equation}
It can be seen from the expressions above that the late time growth of complexity of dyonic black holes are same as that of neutral black holes \cite{Emparan:2021hyr} up to leading order quantum correction due to $\text{CFT}_{3}$, while the charge $q$ dependence is hidden in $M$ and $\mu$.
\subsection{Quantum dressed defects}
In this section, we focus on the $\kappa=1$ case, which can be interpreted as quantum dressed defect geometries, as there is no horizon classically to hide the defect inside. The horizon appears due to quantum backreaction on the geometry $r_{+}\sim\frac{\mu\ell}{2}$ which coincides with $r_{-}$, therefore from the bulk viewpoint this can be interpreted as an extremal Reissner-Nordstr\"om (RN) black hole.

The late time growth of complexity for this case can be found from \eqref{eq:comprate} with $\kappa=1$, the minimum radius in this case is $r_{\text{min}}=\frac{\mu\ell}{2}$. The late time rate of complexity is given by,
\begin{equation}
    \frac{dC_{V}}{d\bar{t}}=\frac{2|M| \ell \mu  }{\ell_{3}}\sqrt{1-\frac{4 q^2}{\mu ^2}}\,,
\end{equation}
In $q\rightarrow0$, this agrees with the neutral case \cite{Emparan:2021hyr}. As the charge is $q=\frac{\mu}{2}$ in extremal limit \cite{Climent:2024nuj}, the complexity growth vanishes. This agrees with the previous observations for a classical 4D Reissner-Nordstr\"om black hole in the extremal limit \cite{Carmi:2017jqz}.


\section{Action complexity}\label{sec:AC}
In this section, we examine the CA proposal for the holographic complexity of the charged quantum corrected black hole. The CA conjecture probes the complexity of the time-evolved thermo-field double state $\psi(t) = \ket{\text{TFD}_{\beta}(t)}$ of the CFT determined from the on-shell gravitational action of the WDW patch
\begin{equation}
{\mathcal{C}}_{A}(\psi)= \frac{I[\mathcal{W}_{\tilde{t}}]}{\pi \hbar} \,.
\end{equation}
The WDW patch is the union of all possible spacelike codimension-one surfaces anchored at constant boundary time $\tilde{t}$. In other words, it is a full causal patch of the Cauchy slice at constant boundary time $\tilde{t}$.  This formulation is sensitive to the global structure of geometry, particularly near singularities and boundaries. Our goal here is to determine whether CA behaves consistently under quantum corrections and accurately reproduces the correct classical limit.

\subsection{Regularized WDW patch}
In general, the action over the WDW patch is diverging; there are two standard ways to regularize it: either to put a cutoff surface near the asymptotic boundary, say at $r=r_{\text{max}}$, or consider the WDW to end on the regularized surface itself \cite{Carmi:2016wjl}. In the first regularization picture, we have a GHY contribution from the cutoff surface and two joint contributions from the joint surfaces created from the intersection of the null boundaries of WDW patch and the cutoff surface, while in the second picture, there is no GHY contribution and we get a single joint contribution from the intersection of the future and past null boundaries of WDW patch. The exact equivalence of these two regularization approaches was shown in \cite{Akhavan:2019zax} by adding counter terms on these cutoff surfaces (timelike boundaries). Also, it has been shown that the UV divergences can be removed completely by adding counter terms on the null boundaries of the WDW patch. Later in \cite{Omidi:2020oit}, the equivalence is shown in the case of a two-sided BTZ black hole truncated by a dynamical timelike ETW brane. Here, we consider the second approach to regularization, in addition to that, rather than considering the full WDW patch till the asymptotic boundary of the $\text{AdS}_{4}$ C-metric at $rx=-\ell$, we clip off the WDW at $r=\infty$, denoted by $\widetilde{W}_{\tilde{t}}$. We have shown explicitly in the appendix \ref{app: fullWDW} that if we consider the full WDW patch, it contributes a constant shift in the total complexity and leaves the late time growth unaffected. To simplify the discussion, we consider $\widetilde{\mathcal{W}}_{\tilde{t}}$ as the relevant object that describes the holographic complexity of the degrees of freedom within the interior of the quBTZ black hole. In particular, we will determine
\begin{equation}
\widetilde{\mathcal{C}}_{A}(\psi)= \frac{I[\widetilde{\mathcal{W}}_{\tilde{t}}]}{\pi \hbar} \,.
\end{equation}
The additivity of the action gives $\mathcal{C}_A = \tilde{\mathcal{C}}_A + \mathcal{C}_{\text{UV}}$, where the UV component is the action evaluated on $\mathcal{W}_{\tilde{t}} \setminus \widetilde{\mathcal{W}}_{\tilde{t}}$. The UV complexity $\mathcal{C}_{\text{UV}}$ corresponds to the short-range correlations in the dCFT system's state. The static nature of the spacetime outside the black hole causes the UV contribution to the action complexity (AC) to become constant at late times. Consequently, the late-time growth of AC is governed by the regularized action $I[\widetilde{\mathcal{W}}_{\tilde{t}}]$.

\subsection{Action computation}
The on-shell action $ I[\widetilde{\mathcal{W}}_{\tilde{t}}] $ of the regularized WDW patch consists of,
\begin{align}\label{eq:total action}
    I[\widetilde{\mathcal{W}}_{\tilde{t}}] = I_{\text{EH}} + I_{\text{Max}} + I_{\text{brane}} +I_{\text{GHY}} +I_{\mu Q}+ I_{\text{Joint}} + I_{\kappa} + I_{\text{ct}} \,,
\end{align}
where
\begin{align}
    \begin{aligned}
    & I_{\mathrm{GHY}} =\frac{1}{8 \pi G_4} \int_{\text {regulator }} d^3 y \sqrt{|h|} K , \qquad
    &I_{\text {Joint }}  & =\frac{1}{8 \pi G_4} \int_{\text {joints }} d^2 y \sqrt{\sigma} \alpha, \\
    & I_\kappa  =\frac{1}{8 \pi G_4} \int_{\mathcal{N}} d \lambda_{n} d^2 y \sqrt{\gamma} \kappa, \qquad
    & I_{\text{ct}} & =\frac{1}{8 \pi G_4} \int_{\mathcal{N}} d \lambda_{n} d^2 y \sqrt{\gamma} \Theta \log \left(L_{c t} \Theta\right).
    \end{aligned}
\end{align}
The total action includes multiple terms: $I_{\mathrm{EH}}$ the Einstein- Hilbert (EH) term with a negative cosmological constant $\Lambda_4 =  - 3/\ell ^2 _4$, $I_{\text{Max}}$ is the Maxwell term, where the integration domain $\mathcal{W}$ is the bulk WDW patch; and $I_{\mathrm{brane}}$ the brane action where $w = \mathcal{W} \cap \mathrm{brane}$ is the intersection between the WDW patch and the brane, with $h$ being the induced metric on the brane as defined in \eqref{eq:actionbulk}. Additionally, $I_{\mathrm{GHY}}$, the Gibbons-Hawking York (GHY) term defined on the regulator surface near AdS boundary, $I_{\mu Q}$ is Maxwell boundary term defined on the regulator surface; $I_{\mathrm{Joint}}$, the joint terms evaluated at the intersections of the null boundaries of the WDW patch with other smooth hypersurfaces (which will be specified below), where $\alpha$ is some function that depends on the `boost angle' between the normals to the hypersurfaces; $I_{\kappa}$, the null boundary terms corresponding to the null boundaries $\mathcal{N}$ of the WDW patch, which would vanish when the null surface is affinely parameterized; $I_{\mathrm{ct}}$, the counter term to make the action invariant under reparameterization of null generators, also defined on $\mathcal{N}$ which could be expressed in terms of its expansion parameter $\Theta=\frac{1}{\sqrt{\gamma}} \frac{\partial\sqrt{\gamma}}{\partial\lambda_{n}}$, $\lambda_{n}$ is an affine parameter and $\gamma$ is induced metric on the null boundaries.
\begin{eqnarray}
    ds^{2}=0,\quad H(r) \dot{t}^2=\frac{ \dot{r}^2}{H(r)}+r^2\left(\frac{ \dot{ x}^2}{G(x)}+G(x) \dot{\phi}^2\right) \,.
\end{eqnarray}
We consider the null boundaries of the WDW patch to be affinely parametrized so that the contribution from $I_{\kappa}$ vanishes, and we have seen that the counter term gives a vanishing contribution to the complexity in the late time regime; the details have been presented in the appendix \ref{app: counter term}. 

Now, to calculate the Action complexity, we proceed to calculate various contributions from the ``regularized" WDW patch to the action.

\subsubsection*{Bulk contribution}
The bulk contribution $I_{\mathrm{bulk}}(\widetilde{\mathcal{W}}_{\tilde{t}})$ to the action \eqref{eq:total action} will consist of the Einstein-Hilbert action $I_{\mathrm{EH}}(\widetilde{\mathcal{W}}_{\tilde{t}})$, Maxwell action $I_{\mathrm{Max}}(\widetilde{\mathcal{W}}_{\tilde{t}})$ and the bare brane action $I_{\mathrm{brane}}(\tilde{w}_{\tilde{t}})$. The \eqref{eq:met-sol} satisfies the local vacuum solution away from the brane, and the presence of the brane affects the curvature via Israel's junction condition as 
\begin{align}
    R-2\Lambda_4 = -\frac{6}{\ell_4^2} + \frac{12 \delta(x)}{\ell r}\,.
\end{align}
The term proportional to $\delta(x)$  collapses on the brane position, and the volume integral becomes an integral over the $w_t$. Thus, the bulk and brane action in the bulk region of the WDW patch is given by,
\begin{align}
     I_{\text{EH}}^{\text{WDW}} & = \frac{1}{16 \pi G_{4}}\left(-\frac{6}{\ell_{4}^{2}}\right)\text{Vol}(\widetilde{\mathcal{W}}_{t})+\frac{1}{16 \pi G_{4}}\left(\frac{12}{\ell}\right)\text{Vol}(\widetilde{w}_{t}) \,,
\end{align}
and the bulk contribution from the Maxwell term is given by
\begin{align}
    I_{\text{Max}} & =    -\frac{1}{4g_{\star}^{2}}\int_{\mathcal{M}} \mathrm{d}^{4} x \sqrt{-g}\left(-\frac{2q^{2}(\ell+r x)^{4}}{\ell^{2}r^{4}}\right) \,.
\end{align}
In addition to that, a bare brane contribution arises from $I_\text{brane}$ in the total bulk contribution. Thus the bulk action becomes
\begin{align}\label{eq:bulkcontri}
    I_{\mathrm{bulk}}(\widetilde{\mathcal{W}}_{\tilde{t}}) =  - \frac{3 \mathrm{Vol}(\widetilde{\mathcal{W}}_{\tilde{t}})}{8 \pi G_4 \ell^2_4} + \frac{\mathrm{Vol}({w}_{\tilde{t}})}{4 \pi G_4 \ell} + I_{\mathrm{Max}}(\widetilde{\mathcal{W}}_{\tilde{t}}) \,,
\end{align}
To compute the bulk action, we first need to determine the volume of the regularized WDW patch, denoted as $\volWt$, along with the brane volume that intersects with the bulk, represented as $\volwt$ , as shown in equation \eqref{eq:bulkcontri}. Following the method outlined in \cite{Emparan:2021hyr}, we can regularize the WDW patch by anchoring the time slice at the boundary $(r = r_{\infty})$. We can then divide the volume integrals into three distinct regions, as illustrated in the figure \ref{fig:penrose1} below.
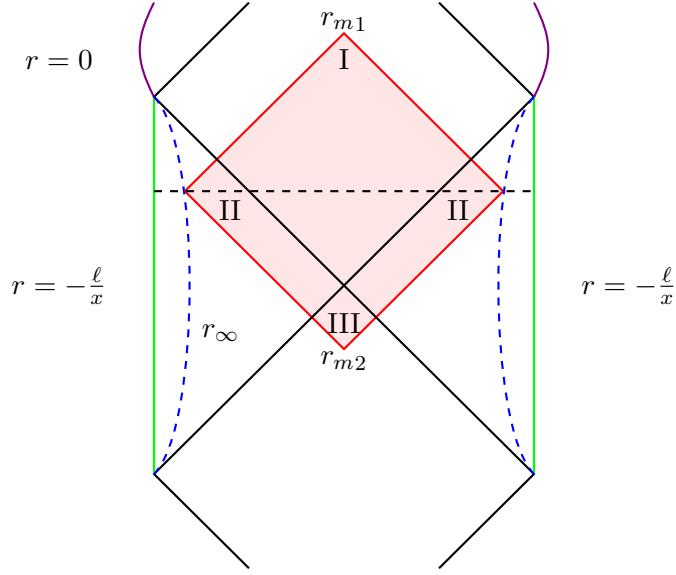
\begin{figure}[ht]
    \centering
    \begin{tikzpicture}[scale=2.5]
    \node[diamond, thick, red, draw, fill=red!10, minimum width = 4.2cm, minimum height = 4.2cm] (d) at (0,0.5) {};
    \draw[thick] (-0.5,-1.5) -- (-1,-1) -- (0,0) -- (1,-1) -- (0.5, - 1.5);
    \draw[green,thick] (-1,1) -- (-1,-1);
    \draw[thick] (-0.5,1.5) -- (-1,1) -- (0,0) -- (1,1) -- (0.5,1.5);
    \draw[green,thick] (1,1) -- (1,-1);
    \draw[dashed,thick] (-1,0.5) -- (1,0.5);

    \draw[dashed, thick,blue] (-1,1) .. controls (-0.75,0.75) and (-0.75,-0.75) .. (-1,-1);
    \draw[thick, violet] (-1,1) .. controls (-1.1,1.2) and (-1.1,1.3) .. (-1.0,1.5);

    \draw[dashed, thick, blue] (1,1) .. controls (0.75,0.75) and (0.75,-0.75) .. (1,-1);
    \draw[thick, violet] (1,1) .. controls (1.1,1.2) and (1.1,1.3) .. (1.0,1.5);

    \node at (0,1.4) {${r_m}_1$};
    \node at (0,1.2) {I};
    \node at (-0.6,0.4) {II};
    \node at (0.6,0.4) {II};
    \node at (0,-0.2) {III};
    \node at (0,-0.4) {${r_m}_2$};
    \node at (-1.5,0) {$r =  - \frac{\ell}{x}$};
    \node at (1.5,0) {$r =  - \frac{\ell}{x}$};
    \node at (-0.65,-0.25) {$r_\infty$};
    \node at (-1.5,1.2) {$r=0$};
    \end{tikzpicture}
    \caption{ Regularized WDW patch}
    \label{fig:penrose1}
\end{figure}
This illustration demonstrates that performing the appropriate integration across the $t$-coordinate for each designated region produces:
\begin{eqnarray}
    \text{Vol}(\widetilde{\mathcal{W}}_{\tilde{t}}^{I})&=&4\int_{0}^{2\pi \Delta}d\phi\int_{0}^{x_{1}}dx\int_{r_{m_{1}}}^{r_{+}} dr\frac{\ell^{4}r^{2}}{(\ell+x r)^{4}}(\tilde{t}-r_{*}(r))\,,\nonumber\\
    \text{Vol}(\widetilde{\mathcal{W}}_{\tilde{t}}^{II})&=&4\int_{0}^{2\pi \Delta}d\phi\int_{0}^{x_{1}}dx\int_{r_{+}}^{r_{\infty}} dr\frac{\ell^{4}r^{2}}{(\ell+x r)^{4}}(2r_{*}(r))\,,\nonumber\\
    \text{Vol}(\widetilde{\mathcal{W}}_{\tilde{t}}^{III})&=&4\int_{0}^{2\pi \Delta}d\phi\int_{0}^{x_{1}}dx\int_{r_{m_{2}}}^{r_{+}} dr\frac{\ell^{4}r^{2}}{(\ell+x r)^{4}}(-\tilde{t}+r_{*}(r))\,,
\end{eqnarray}
where, $r_{*}$ is the tortoise coordinate, defined as,
\begin{equation}
    r_{*}(r)=\int_{r}^{\infty}\frac{dr'}{H(r')}.
\end{equation}
The factors of 2 are included to account for the two-sided spacetime (for $ x > 0$ and $x<0$), and reflection symmetry in $t$. Thus, the total contribution of $\Wtt$ from the Einstein-Hilbert part is 
\begin{equation}
    \text{Vol}(\widetilde{\mathcal{W}}_{\tilde{t}})=V_{0}+8 \pi \Delta \ell^{4} \int_{0}^{x_{1}}dx \int_{r_{m_{1}}}^{r_{m_{2}}}dr\frac{r^{2}}{(\ell +x r)^{4}}(\tilde{t}-r_{*}(r))\,,
\end{equation}
where the time-independent volume part is given by,
\begin{equation}
    V_{0}=16 \pi \Delta \ell^{4}\int_{0}^{x_{1}}dx\int_{r_{+}}^{r_{\infty}}dr\frac{r^{2}}{(l +x r)^{4}}r_{*}(r)\,.
\end{equation}
Similarly, repeating the previous analysis for the volume on the brane \emph{i.e.}, the volume of $w_{\tilde{t}}$ can be found as:
\begin{eqnarray}
  \text{Vol}(w_{\tilde{t}})&=&\text{v}_{0}+4 \pi \Delta \int_{r_{m_{1}}}^{r_{m_{2}}}dr (r(\tilde{t}-r_{*}(r))) , \\
   \text{v}_{0}&=&8 \pi \Delta\int_{r_{+}}^{r_{\infty}}dr (r r_{*}(r)) .
   \end{eqnarray}
The last contribution to the $ I_{\mathrm{bulk}}(\widetilde{\mathcal{W}}_t)$ comes from the Maxwell part of the action on the WDW patch
\begin{eqnarray}
   \text{Vol}(\mathcal{\widetilde{W}}^{\text{F}}_{\tilde{t}})&=&V^{\text{F}}_{0}-16 q \ell^{2}\pi \Delta \int_{0}^{x_{1}}dx\int_{r_{m_{1}}}^{r_{m_{2}}}dr\frac{1}{r^{2}}(\tilde{t}-r_{*}(r)) \,,\\
   V^{\text{F}}_{0}&=&-32 q \ell^{2}\pi \Delta \int_{0}^{x_{1}}dx\int_{r_{+}}^{r_{max}}dr\frac{1}{r^{2}}(r_{*}(r))\,.
\end{eqnarray}

\subsubsection*{GHY contribution}
The GHY term will contribute only when we consider the full WDW patch with a cutoff surface near the asymptotic boundary, rather than the regularized WDW patch, which ends at the cutoff surface. Also, unlike the neutral case, the regularized WDW patch does not touch the singularity, so there is no need to put a cutoff surface near the singularity, and there is no GHY-like contribution from the singularity.

\subsubsection*{Maxwell boundary contribution}
As we anchor the WDW patch at the regulator surface itself, there is no regulator surface from which we get a Maxwell boundary contribution. If we, rather than anchoring the WDW patch at the regulator surface, introduce a cutoff surface, then we get a non-zero Maxwell boundary contribution. More details on this have been given in section \ref{sub: Maxwell boundary}.
\subsubsection*{Joint contributions}
The joint action  is given by
\begin{align}
I_{\text{Joint}} = \frac{1}{8 \pi G_4} \int d^2 y \sqrt{\sigma} a \, \quad \text{with} \quad  a = \log \left(- n^{(L)}_{\mu} n^{(R)}{}^{\mu} / 2 \right).
\end{align}
We need to consider the contribution from joint 1 at $r_m^1$ and joint 2 at $r_m^2$ (Intersection of null boundaries in the black hole interior region \emph{i.e.} $r_- < r_m^i < r_+$ for $i \in (1,2)$ respectively). There is also a third contribution from joint 3 at $r=r_{\infty}$.
\begin{figure}[ht]
    \centering
    \begin{tikzpicture}[scale=2.5]
    \node[diamond, thick, red, draw, fill=red!10, minimum width = 4.2cm, minimum height = 4.2cm] (d) at (0,0.4) {};
    \draw[thick] (-0.75,-1.25) -- (-1,-1) -- (0,0) -- (1,-1) -- (0.75, - 1.25);
    \draw[green,thick] (-1,1) -- (-1,-1);
    \draw[thick] (-0.5,1.5) -- (-1,1) -- (0,0) -- (1,1) -- (0.5,1.5);
    \draw[green,thick] (1,1) -- (1,-1);

    \draw[dashed, thick,blue] (-1,1) .. controls (-0.75,0.75) and (-0.75,-0.75) .. (-1,-1);
    \draw[thick, violet] (-1,1) .. controls (-1.1,1.2) and (-1.1,1.3) .. (-1.0,1.5);

    \draw[dashed, thick, blue] (1,1) .. controls (0.75,0.75) and (0.75,-0.75) .. (1,-1);
    \draw[thick, violet] (1,1) .. controls (1.1,1.2) and (1.1,1.3) .. (1.0,1.5);

    \filldraw[violet] (0.82,0.4) circle (1pt);
    \filldraw[violet] (-0.82,0.4) circle (1pt);
    \filldraw[violet] (0,1.22) circle (1pt);
    \filldraw[violet] (0,-0.42) circle (1pt);
    
    \node[text=violet] at (0.70,0.4) {$J_3$};
    \node[text=violet] at (-0.70,0.4) {$J_3$};
    \node[text=violet] at (0,1.1) {$J_1$};
    \node[text=violet] at (0,-0.3) {$J_2$};
    \end{tikzpicture}
    \hspace{1cm}
    \begin{tikzpicture}[scale=2.5]
    \node[diamond, thick, red, draw, fill=red!10, minimum width = 4.2cm, minimum height = 4.2cm] (d) at (0,0.4) {};
    \draw[thick] (-0.75,-1.25) -- (-1,-1) -- (0,0) -- (1,-1) -- (0.75, - 1.25);
    \draw[green,thick] (-1,1) -- (-1,-1);
    \draw[thick] (-0.5,1.5) -- (-1,1) -- (0,0) -- (1,1) -- (0.5,1.5);
    \draw[green,thick] (1,1) -- (1,-1);

    \draw[dashed, thick,blue] (-1,1) .. controls (-0.75,0.75) and (-0.75,-0.75) .. (-1,-1);
    \draw[thick, violet] (-1,1) .. controls (-1.1,1.2) and (-1.1,1.3) .. (-1.0,1.5);

    \draw[dashed, thick, blue] (1,1) .. controls (0.75,0.75) and (0.75,-0.75) .. (1,-1);
    \draw[thick, violet] (1,1) .. controls (1.1,1.2) and (1.1,1.3) .. (1.0,1.5);
    \draw[thick, cyan] (0.83,0.4) -- (0,1.23) -- (-0.83,0.4) -- (0,-0.44) -- (0.83, 0.4);
    \node[text=violet] at (0.5,1) {$\mathcal{B}_1$};
    \node[text=violet] at (0.5,-0.2) {$\mathcal{B}_2$};
    \node[text=violet] at (-0.5,1) {$\mathcal{B}_1$};
    \node[text=violet] at (-0.5,-0.2) {$\mathcal{B}_2$};
    \end{tikzpicture}
    \caption{\emph{ Left}: The violet points are `standard joint surfaces' in the bulk for a constant $(x,\phi)$ section of the geometry.  \emph{Right:} The sky blue lines are `new joint surfaces' for a constant $\phi$ section of the brane. The sky blue lines represent two-dimensional surfaces that extend in the $\phi$ direction along the brane.}
    \label{fig:penrosejoint}
\end{figure}
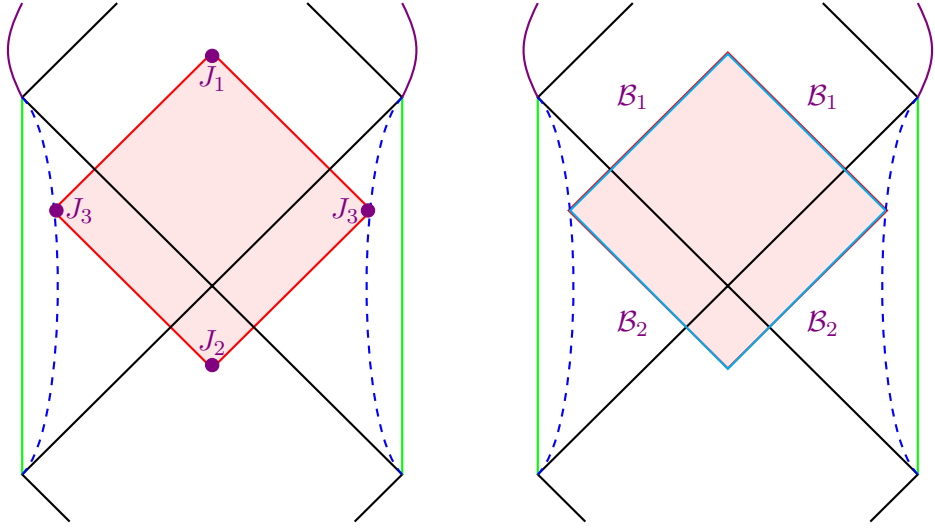
The contributions from joint 1 and joint 2 are given by,
\begin{eqnarray}
I_{\text{Joint}}^{J_{i}}=\frac{2}{8\pi G_{4}}\int_{0}^{x_{1}}dx\int_{0}^{2\pi \Delta}d\phi\left(\frac{\ell^{2}r^{2}}{(\ell+r x)^{2}}\log\left(\frac{\alpha_{1}\alpha_{2}(\ell+r x)^{2}}{\ell^{2}H(r)}\right)\right)_{r=r_{m_{i}}} \,,
\end{eqnarray}
The contribution from the 3rd joint
\begin{eqnarray}
I_{\text{Joint}}^{J_{3}}=-\frac{2}{8\pi G_{4}}\int_{0}^{x_{1}}dx\int_{0}^{2\pi \Delta}d\phi\left(\frac{\ell^{2}r^{2}}{(\ell+r x)^{2}}\log\left(\frac{\alpha_{1}\alpha_{2}(\ell+r x)^{2}}{\ell^{2}H(r)}\right)\right)_{r=r_{\infty}}\,.
\end{eqnarray}
As the bulk of WDW patch also include a brane, there is an additional contribution to the joint action from joint surfaces created by joining the two sides of the WDW patch at the brane (at $x=0$). The contribution from these surfaces, denoted by $\mathcal{B}_{1}$ and $\mathcal{B}_{2}$, shown in figure \ref{fig:penrosejoint}, is given by
\begin{eqnarray}
I_{\text{Joint}}^{\mathcal{B}_{1,2}}=\int a dt d\phi , \qquad \qquad a=\log\left(-\frac{n^{(+)}.n^{(-)}}{2}\right)\,,
\end{eqnarray}
where, $n^{(\pm)}_{\mu}$ are normal vectors on the boundaries of the WDW patch in $x>0$ and $x<0$ regions, respectively\footnote{given by, $n_{\mu}=dt\pm \frac{dr}{H(r)}$ for $\mathcal{B}_{1}$ and $\mathcal{B}_{2}$ respectively.}. As the normal vectors $n^{(+)}$ and $n^{(-)}$ are continuous across the brane \emph{i.e.} $n^{(+)}=n^{(-)}$. This leads to the vanishing contribution to the joint action as the boost vector vanishes $a=0$.

\subsection{Late time growth}
As in late time regime $r_{m_{1}}=r_{-}$ and  $r_{m_{2}}=r_{+}$
and,
\begin{equation}
    \frac{\partial r_{{m}_{1}}}{\partial \tilde{t}}= H(r), \quad \frac{\partial r_{{m}_{2}}}{\partial \tilde{t}}=- H(r),
\end{equation}
The joint term in the late time regime simplifies to:
\begin{eqnarray}
I_{\text{Joint}}^{J_{1}}&=&\frac{\Delta}{2 G_{4}}\int_{0}^{x_{1}}dx\left(\frac{\ell^{2}r^{2}}{(\ell+r x)^{2}}\log\left(\frac{\alpha_{1}\alpha_{2}(\ell+r x^{2})}{\ell^{2}H(r)}\right)\right)_{r=r_{-}}\,,\nonumber\\
I_{\text{Joint}}^{J_{2}}&=&\frac{\Delta}{2 G_{4}}\int_{0}^{x_{1}}dx\left(\frac{\ell^{2}r^{2}}{(\ell+r x)^{2}}\log\left(\frac{\alpha_{1}\alpha_{2}(\ell+r x)^{2}}{\ell^{2}H(r)}\right)\right)_{r=r_{+}}\,,
\end{eqnarray}
and the growth in late time\footnote{we consider the growth in regularized time $\bar{t}=\Delta \tilde{t}$.},
\begin{eqnarray}
    \frac{dI_{\text{Joint}}}{d\bar{t}}=\frac{\Delta ^2 \ell x_1}{2 G_4 } \left(\frac{r_+^2 \left(-\frac{2 \ell^2 q^2}{r_+^3}+\frac{\ell \mu }{r_+^2}+\frac{2 r_+}{\ell_{3}^2}\right)}{\ell+r_+ x_1}+\frac{r_-^2 \left(-\frac{2 \ell^2 q^2}{r_-^3}+\frac{\ell \mu }{r_-^2}+\frac{2 r_-}{\ell_{3}^2}\right)}{\ell+r_- x_1}\right)\,.
\end{eqnarray}
Total joint contribution to the late time growth can be rewritten as,
    \begin{equation}
            \frac{dI_{\text{Joint}}}{d\bar{t}}=2TS_{\text{gen}}\bigg|^{r_+}_{r_-}\,.
    \end{equation}
Also, the contribution of bulk terms in the late-time growth is given by
\begin{eqnarray}
   \frac{dC_{\text{bulk}}}{d\bar{t}}&=&\frac{1}{\pi} \left. \left(\frac{\Delta ^2 r^2}{2 G_{4} \ell} -\frac{16 \pi  \Delta ^2 \ell^2 x_{1} \left(e^2-g^2\right)}{g_{\star}^2 \ell_{\star}^2 r}-\frac{\Delta ^2 \ell r^3 x_{1} (2 \ell+r x_{1})}{2 G_{4} \ell_{4}^2 (\ell+r x_{1})^2}\right) \right|^{r=r_+} _{r=r_-}.
\end{eqnarray}
The total late time growth of complexity can be rewritten in terms of thermodynamic quantities as,
\begin{eqnarray}
    \frac{dC_{A}}{d\bar{t}}&=&\frac{dC_{\text{bulk}}}{d\bar{t}}+\frac{dC_{\text{Joint}}}{d\bar{t}}\nonumber\\
    &=&\frac{1}{\pi}\Bigg(\frac{2 \left(\nu ^2-\nu ^4 \gamma ^2 z^4-\nu ^3 \gamma ^2 z^3-\nu ^3 z^3+\nu  z+1\right)}{\nu ^2 \left(\nu ^2 \gamma ^2 z^4+\nu  \gamma ^2 z^3-\nu  z^3+1\right)}M-2 TS_{\text{gen}}\nonumber\\&&-2Q_{e}\Phi_{e}+2V_{4}P_{4}-2\tau A_{\tau}\Bigg)\Bigg|^{{r_{+}}}_{r_{-}}\,,
\end{eqnarray}
\begin{figure}
        \centering
        \includegraphics[width=0.7\linewidth]{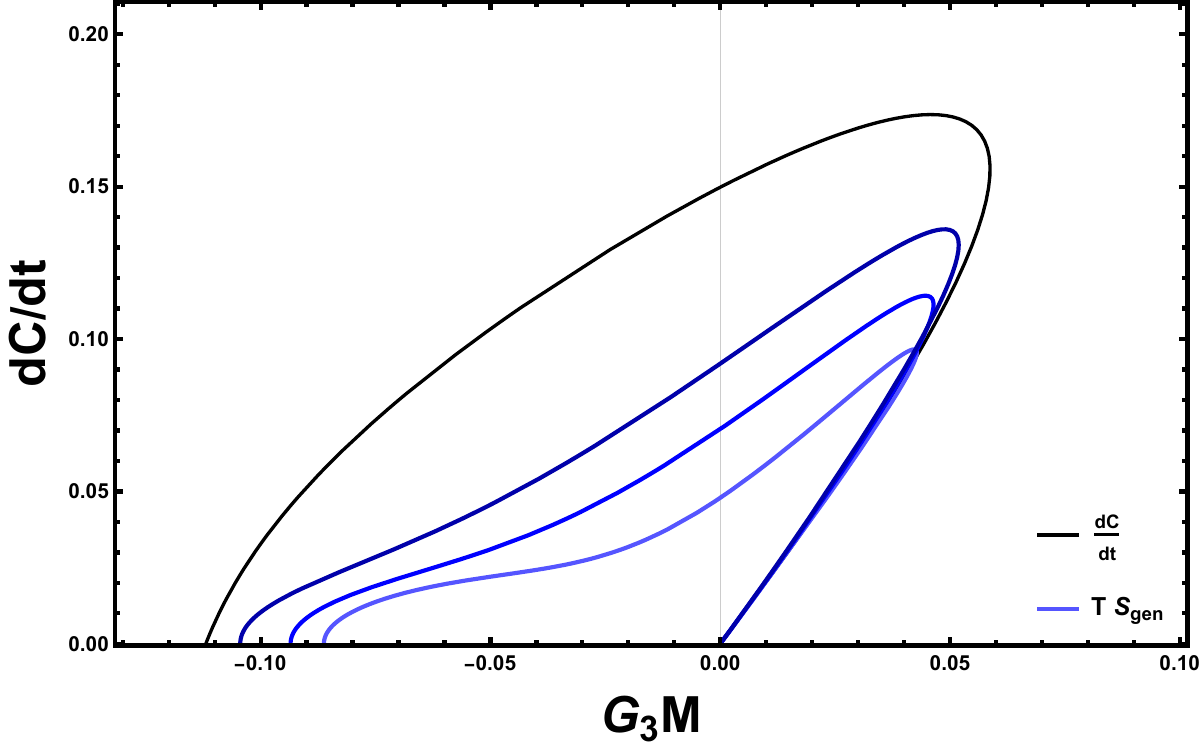}
        \caption{ The late-time slope of AC, in black, is independent of $\ell$}
        \label{fig:ComplexityvsG3M}
\end{figure}
We also present the dependence of late time complexity growth on $G_{3}M$ in figure \ref{fig:ComplexityvsG3M}, which shows that for small values of $G_{3}M$ \emph{i.e} for small masses the complexity growth agrees with $TS_{\text{gen}}$.\\

\noindent For $\ell\rightarrow0$ limit, $r_{\pm}$ is found using $H(r_{\pm})=0$,
\begin{eqnarray}
    r_{+}&=& \ell_{3}+\frac{\mu  \ell}{2},\\
        r_{-}&=&- \ell_{3}+\frac{\mu  \ell}{2}.
\end{eqnarray}
An order-by-order analysis of the complexity growth shows that, up to $\mathcal{O}(\ell)$ it satisfies
    \begin{eqnarray}
        \frac{d\mathcal{C}_{{A}}}{d\bar{t}}&=&\frac{1}{\pi}\left(2\frac{1+x_{1}^{2}}{x_{1}^{2}}M-2Q_{e}\Phi^{e}\right)\Bigg|_{r=r_{-}}^{r=r_{+}}\nonumber\\
        &=&\frac{1}{\pi}\left(2(\mu x_{1}+q^{2}x_{1}^{2})M-2Q_{e}\Phi^{e}\right)\Big|_{r=r_{-}}^{r=r_{+}}\,.
    \end{eqnarray}
 It can be noticed that the late time growth of complexity for charged quantum black holes attained by following the CA conjecture agrees with the neutral case \cite{Emparan:2021hyr} in $q\rightarrow0$ limit up to a factor\footnote{Due to the absence of a GHY contribution from the singularity which gives a non-vanishing contribution in the neutral quantum black hole case.}. A significant feature of this result is the prefactor of the mass $M$ term, which is dependent on the parameter $x_{1}$; this distinct it from the previous observations for classical black holes \cite{Reynolds:2016rvl, Brown:2015bva} that shows the agreement between CV and CA results up to a prefactor dependent on the spacetime dimension. Another significant point to notice is that the rate of complexity vanishes for a purely magnetic charged case. This point has been previously discussed in \cite{Goto:2018iay}; the details of which are given in the section below.
\subsection{Effects of Maxwell boundary term}\label{sub: Maxwell boundary}
In \cite{Goto:2018iay}, it has been shown that for charged black holes, a boundary term for spacelike/timelike boundaries can be added in the action contribution to the complexity.
\begin{equation}
    I_{\mu Q}  =\frac{\gamma}{g_{\star}^{2}}\int d\Sigma_{\mu}F^{\mu\nu}A_{\nu}\,.
\end{equation}
This boundary term can be transformed into a bulk term on shell and make a final contribution to the complexity growth by modifying the coefficient of the contribution from the bulk Maxwell term. 
\begin{equation}
    I_{\text{Max}}+I_{\mu Q}  =\frac{(2\gamma-1)}{2g_{\star}^{2}}\int d^{4}x~\sqrt{-g}~F^{\mu\nu}F_{\mu\nu}\,,
\end{equation}
The addition of this term into the action resolved the discrepancy observed in the complexity of the Reissner - Nordstr\"om black hole; therefore, that of JT gravity obtained from the dimensional reduction of the extremal RN solution\cite{Goto:2018iay}. It has been observed that different choice of the coefficient $\gamma$ has different effects. For example, if we choose $\gamma=\frac{1}{2}$, then there is no contribution from the bulk Maxwell term, and the final complexity has electric and magnetic charge on equal footing. Without this surface term $\gamma=0$, the final late time complexity growth of purely magnetically charged black holes vanishes while the electrically charged black holes have a non-vanishing complexity, and this effect is reversed (electric $\longleftrightarrow$ magnetic) for $\gamma=1$. Also, the contribution of this term only affects the IR regime, in the UV regime (near the boundary) it gives a vanishing contribution. In our setup, such a contribution can only appear if we introduce a cutoff (at $r_{\infty}$) near the asymptotic boundary, and anchor the WDW patch at aymptotic boundary rather than on the cutoff surface itself\footnote{This will also give three additional contributions: GHY contribution from the cutoff surface, and two joint terms. But they will not give any contribution to the late time growth of complexity.}. This will change the coefficient of Maxwell bulk term. As our final result also reflects that the late-time growth of a purely magnetic charged solution vanishes. We proceed in this way, and observe the effect of Maxwell boundary term. This makes a contribution to the late-time growth of complexity as,
\begin{eqnarray}
    \frac{dI_{\mu Q}}{d\bar{t}}=\frac{2 \gamma  \Delta ^2 \ell^2 x_1 \left(g^2-e^2\right)}{G_4 r}\Bigg|^{r_{+}}_{r_{-}},
\end{eqnarray}
which for $\gamma=\frac{1}{2}$ modifies the late time complexity growth as,
\begin{eqnarray}
    \frac{dC_{A}}{d\bar{t}}&=&\frac{dC_{\text{bulk}}}{d\bar{t}}+\frac{dC_{\text{Joint}}}{d\bar{t}}\nonumber\,,\\
    &=&\frac{1}{\pi}\Bigg(\frac{2 \left(\nu ^2-\nu ^4 \gamma ^2 z^4-\nu ^3 \gamma ^2 z^3-\nu ^3 z^3+\nu  z+1\right)}{\nu ^2 \left(\nu ^2 \gamma ^2 z^4+\nu  \gamma ^2 z^3-\nu  z^3+1\right)}M-2 TS_{\text{gen}}\nonumber\\&&-Q_{e}\Phi_{e}-Q_{g}\Phi_{g}+2V_{4}P_{4}-2\tau A_{\tau}\Bigg)\Bigg|^{{r_{+}}}_{r_{-}}\,.
\end{eqnarray}
An order-by-order analysis of the complexity growth shows that, upto $\mathcal{O}(\ell)$ it satisfies
    \begin{eqnarray}
        \frac{d\mathcal{C}_{A}}{d\bar{t}}&=&\frac{1}{\pi}\left(2\frac{1+x_{1}^{2}}{x_{1}^{2}}M-Q_{e}\Phi_{e}-Q_{g}\Phi_{g}\right)\Bigg|_{r=r_{-}}^{r=r_{+}}\nonumber\\
        &=&\frac{1}{\pi}\left(2(\mu x_{1}+q^{2}x_{1}^{2})M-Q_{e}\Phi_{e}-Q_{g}\Phi_{g}\right)\Big|_{r=r_{-}}^{r=r_{+}}\,.
    \end{eqnarray}
 Also, it can be noticed that for $x_{1}$ large, the known Lloyd bound for dyonic black holes \cite{Ovgun:2018jbm} is satisfied.
\subsection{Quantum dressed defects}
For the case of $\kappa=1$ \emph{i.e.} for quantum dressed charged conical defects in $\text{AdS}_{3}$, an order-by-order expansion of the horizon radius found from $H(r)=0$, gives
\begin{eqnarray}
    r_{\pm}=\frac{\mu\ell}{2} \,.
\end{eqnarray}
Therefore, the late time growth of complexity in this case gives vanishing results up to $\mathcal{O}(\ell)$, which agrees with the CV as well as previous classical observations \cite{Carmi:2017jqz}. 


\section{Discussion}\label{sec:discussion}
In this paper, we have studied the holographic entanglement entropy and holographic complexity of a three-dimensional dyonic quantum black hole as well as quantum dressed dyonic conical defects in $\text{AdS}_{3}$ using the setup of braneworld holography. We have computed the entanglement entropy of $\text{CFT}_{3}$ with a dyonic charged defect whose bulk dual is described by an $\text{AdS}_{4}$-C spacetime with a $\text{AdS}_{3}$ brane, which extends to the position of the defect on the boundary. This has been achieved numerically following the RT prescription for both connected and disconnected (island) RT surfaces up to all orders of quantum corrections due to the bulk quantum fields. We have shown the dependence of entanglement entropy on the subregion size for the connected RT surface, in figure \ref{fig:Avsepsilon-1}, for both charged and neutral cases. It can be seen from the plots that for both quantum black holes and quantum dressed defects the entanglement entropy saturates after an initial increment as the subregion size increases. For the disconnected RT surface, we have studied the behavior of EE with the island region (the intersecting point $r_{0}$ on the brane), within this we have seen that there is an extremum point which minimizes the EE, as shown in figure \ref{fig:Avsr0k-1} and \ref{fig:Avsr0k1} for both quantum black holes and quantum dressed defects in $\text{AdS}_{3}$, which indicates towards the validity of island prescription in this setup.

To study the holographic complexity, we followed both volume complexity and action complexity conjectures. Using the volume complexity conjecture, we could only proceed order by order in the backreaction parameter $\ell$ or $g_{\text{eff}}$, as the extremization of the generic volume functional gets very challenging. Therefore, we restrict ourselves to leading $\mathcal{O}(\ell)$ quantum corrections. As the explicit dependence of charge $q$ appears only at $\mathcal{O}(\ell^{2})$ we do not see any explicit dependence on the charge parameter $q$ in this leading quantum correction; it appears only through its effect on the mass $M$ and the real parameter $\mu$. The late time complexity growth for dyonic quantum black holes following CV is given by,
\begin{equation}
\frac{dC_{V}}{d\bar{t}}=2M+2\sqrt{2}\frac{\mu\ell}{\ell_{3}}M  \,.  
\end{equation}
The leading order piece is that of a classical BTZ black hole, and the first quantum corrected piece is the same as the previously observed neutral and rotating quantum blackholes. This represents the universal nature of quantum corrections to the holographic complexity. It has already been observed in \cite{Emparan:2021hyr} and \cite{Chen:2023tpi} that the leading order quantum correction to the late time growth of complexity deviates from the expected behavior $TS_{\text{gen}}$. Furthermore, for the quantum dressed dyonic defects, we found a vanishing late time complexity growth, which is consistent, as from the four-dimensional bulk perspective it represents a four-dimensional extremal Reissner-Nordstr\"om black holes \cite{Carmi:2017jqz}.

In the study of holographic complexity of quantum black holes, the action conjecture turns out to be much more feasible compared to the volume conjecture. It enabled us to find the holographic complexity, which is exact up to all orders in backreaction. We have studied the fully quantum corrected late-time growth of the complexity following AC, and expressed it in terms of thermodynamic quantities.
\begin{eqnarray}
    \frac{dC_{A}}{d\bar{t}}&=&\Bigg(\frac{2 \left(\nu ^2-\nu ^4 \gamma ^2 z^4-\nu ^3 \gamma ^2 z^3-\nu ^3 z^3+\nu  z+1\right)}{\nu ^2 \left(\nu ^2 \gamma ^2 z^4+\nu  \gamma ^2 z^3-\nu  z^3+1\right)}M-2 TS_{\text{gen}}\nonumber\\&&-Q_{e}\Phi_{e}-Q_{g}\Phi_{g}+2V_{4}P_{4}-2\tau A_{\tau}\,\Bigg)\Bigg|^{r=r_{+}}_{r=r_{-}}.
\end{eqnarray}
This result is closely related to the proposed Lloyd bound for dyonic black holes \cite{Ovgun:2018jbm}, but with an extended set of thermodynamic quantities. There is a recent work \cite{Aguilar-Gutierrez:2023ccv} which presents a detailed study of the validity of the Lloyd bound in the braneworld setup of $\text{AdS}_{3}$ with JT gravity localized on the brane. We aim to conduct such a detailed analysis with some generalization added to our framework in the near future. 

Furthermore, the CA and CV results can be distinguished order by order in the backreaction parameter $\ell$. In the case of classical black holes, the late-time growth usually agrees up to a dimension-dependent constant factor. Here, however, it instead appears as a constant factor determined by $x_{1}$, which sets the upper bound of the boundary coordinate $x$.

Our findings show that dyonic quantum black hole geometries encode quantum corrections in entanglement entropy and holographic complexity in a universal manner, consistent with and extending earlier studies \cite{Emparan:2021hyr, Chen:2023tpi}. In particular, the behavior we observe suggests a deeper unifying structure that links quantum-corrected spacetime geometry with information-theoretic quantities, thereby strengthening the connection between black hole physics and quantum information.


This work also motivates multiple extensions that we intend to approach in the future. One straightforward extension of our study is to include rotation and U(1) charge, as discussed in \cite{Bhattacharya:2025tdn}. The complexity of the quantum rotating BTZ black hole has already been explored in \cite{Chen:2023tpi}. A deeper investigation into entanglement and complexity from an effective field theory perspective also remains an interesting problem \cite{Couch:2016exn, Belin:2021bga, Jorstad:2023kmq}. 

Another natural modification involves replacing the bath, which, in our study, has played a subsidiary role, with an alternative gravitating system by incorporating a second brane into the setup. Following this approach will allow us to extend our analysis to wedge holography \cite{Mollabashi:2014qfa}. Moreover, we can extend to dynamical braneworld scenarios, such as DGP brane constructions \cite{Dvali:2000hr}. Another interesting feature is the study of the first law of complexity in the case of quantum black hole, which states that the variation of complexity depends only on the end point of the trajectory \emph{i.e.} target state \cite{Bernamonti:2019zyy}. These extensions could provide new insights into how entanglement and complexity bounds behave within broader holographic frameworks.

\acknowledgments
We want to thank Subho R Roy for the various discussions and comments. We would like to express our gratitude to Suvankar Dutta, Nabamita Banerjee, Taniya Mandal, Arpita Mitra, and Arpan Bhattacharyya for their valuable comments on the manuscript. We also appreciate Shinji Hirano for his technical feedback on this work. We sincerely acknowledge the initiative of the ``$\text{ST}^4 - 2024$ workshop'', the platform where the stage of this work is laid out. G.S.P. wishes to extend thanks to Imtak Jeon and Robert de Mello Koch for their hospitality in Huzhou. G.S.P. also expresses appreciation to Subho R. Roy for organizing the visit to IIT Hyderabad during the final stages of the project. The work of S.P. is supported by the ANRF-SERB research project CRG/2023/001120, ``\emph{Many Facets of Complexity: From Chaos to Thermalization}''. The work of G.S.P. is supported by the National Natural Science Foundation of China (NSFC) under Grant No. 12247103.


\appendix

\section{Counter term contribution:}\label{app: counter term}
The counter term/LMPS term contribution is necessary to take into account to get rid of the ambiguity by reparameterization of the null generator parameter $\lambda_{n}$ \cite{Lehner:2016vdi}. The counter term is given by,
\begin{eqnarray}
    I_{\text{ct}}=\frac{1}{8\pi G_{4}}\int_{\mathcal{N}_{i}}d\lambda_{n}~ \int d\phi ~\int dx\sqrt{\gamma}\Theta \ln(L_{\text{ct}}|\Theta|)
\end{eqnarray}
We choose $L_{\text{ct}}=1$, the induced metric on the null boundaries $dt=\pm\frac{dr}{H(r)}$ is given by,
\begin{eqnarray}
    ds^{2}=\frac{\ell^{2}}{(\ell+r x)^{2}}r^{2}\left(\frac{dx^{2}}{G(x)}+G(x)d\phi^{2}\right)
\end{eqnarray}
We choose the parameter $\lambda_{n}$ such that $\kappa$ vanishes \emph{i.e.}
\begin{eqnarray}
    k^{\mu}\nabla_{\mu}k^{\nu}=0,\quad k^{\mu}=\frac{\partial x^{\mu}}{\partial\lambda}
\end{eqnarray}
from which we find the form of $\lambda_{n}(r)$. Then the counter term is given by,
\begin{eqnarray}
    I_{\text{ct}}=\frac{1}{4 G_{4}}\int_{\mathcal{N}_{i}}dr~ \lambda_{n}'(r)  ~\int dx\frac{\ell^{2}r^{2}}{(\ell+rx)^{2}}\Theta \ln(|\Theta|)
\end{eqnarray}
It can be seen from the above expression that the contribution from the counter term in complexity growth is
\begin{eqnarray}
    \frac{dI_{\text{ct}}}{dt}&\sim&\frac{dr}{dt}\nonumber\\
    \frac{dI_{\text{ct}}}{dt}&\sim& H(r)\nonumber\\
\end{eqnarray}
which in late time vanishes as $H(r_{\pm})=0$
\section{Full WDW patch}\label{app: fullWDW}

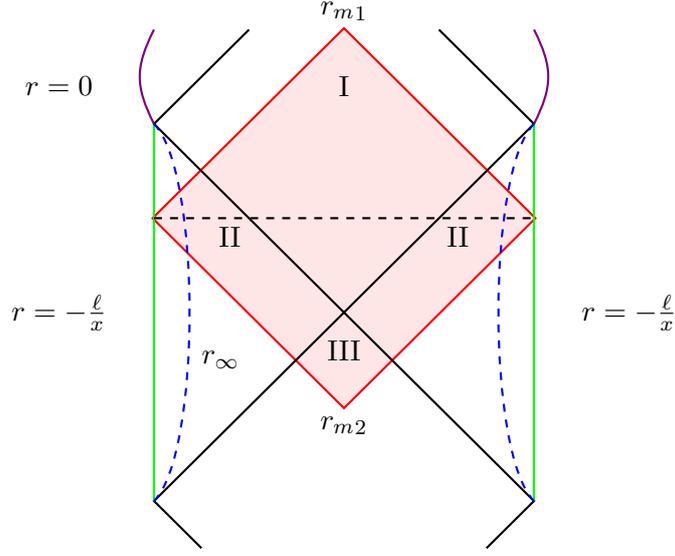
\begin{figure}[ht]
    \centering
    \begin{tikzpicture}[scale=2.5]

    \node[diamond, thick, red, draw, fill=red!10, minimum width = 5.05cm, minimum height = 5.05cm] (d) at (0,0.5) {};
    \draw[thick] (-0.75,-1.25) -- (-1,-1) -- (0,0) -- (1,-1) -- (0.75, - 1.25);
    \draw[green,thick] (-1,1) -- (-1,-1);
    \draw[thick] (-0.5,1.5) -- (-1,1) -- (0,0) -- (1,1) -- (0.5,1.5);
    \draw[green,thick] (1,1) -- (1,-1);
    \draw[dashed,thick] (-1,0.5) -- (1,0.5);

    \draw[dashed, thick,blue] (-1,1) .. controls (-0.75,0.75) and (-0.75,-0.75) .. (-1,-1);
    \draw[thick, violet] (-1,1) .. controls (-1.1,1.2) and (-1.1,1.3) .. (-1.0,1.5);

    \draw[dashed, thick, blue] (1,1) .. controls (0.75,0.75) and (0.75,-0.75) .. (1,-1);
    \draw[thick, violet] (1,1) .. controls (1.1,1.2) and (1.1,1.3) .. (1.0,1.5);

    \node at (0,1.6) {${r_m}_1$};
    \node at (0,1.2) {I};
    \node at (-0.6,0.4) {II};
    \node at (0.6,0.4) {II};
    \node at (0,-0.2) {III};
    \node at (0,-0.6) {${r_m}_2$};
    \node at (-1.5,0) {$r =  - \frac{\ell}{x}$};
    \node at (1.5,0) {$r =  - \frac{\ell}{x}$};
    \node at (-0.65,-0.25) {$r_\infty$};
    \node at (-1.5,1.2) {$r=0$};
    \end{tikzpicture}
    \caption{Full WDW patch}
    \label{fig:penrose2}
\end{figure}

\textbf{Bulk contribution:}

\begin{eqnarray}
    \text{Vol}(\mathcal{W}_{t}^{I})&=&4\int_{0}^{2\pi \Delta}d\phi\int_{0}^{x_{1}}dx\int_{r_{m_{1}}}^{r_{+}} dr\frac{\ell^{4}r^{2}}{(\ell+x r)^{2}}(t+r_{*}(r_{b})-r_{*}(r))\nonumber\\
    \text{Vol}(\mathcal{W}_{t}^{II})&=&4\int_{0}^{2\pi \Delta}d\phi\int_{0}^{x_{1}}dx\int_{r_{+}}^{r_{max}} dr\frac{\ell^{4}r^{2}}{(\ell+x r)^{2}}(2r_{*}(r)-2r_{*}(r_{b}))\nonumber\\
    \text{Vol}(\mathcal{W}_{t}^{III})&=&4\int_{0}^{2\pi \Delta}d\phi\int_{0}^{x_{1}}dx\int_{r_{m_{2}}}^{r_{+}} dr\frac{\ell^{4}r^{2}}{(\ell+x r)^{2}}(-t-r_{*}(r_{b})+r_{*}(r))
\end{eqnarray}

\begin{equation}
    \text{Vol}(\mathcal{W}_{t})=V_{0}+8 \pi \Delta \ell^{4}\int_{0}^{x_{1}}dx\int_{r_{m_{1}}}^{r_{m_{2}}}dr\frac{r^{2}}{(\ell +x r)^{2}}(t+r_{*}(r_{b})-r_{*}(r))
\end{equation}
where the time-independent volume part is given by,
\begin{equation}
    V_{0}=16 \pi \Delta \ell^{4}\int_{0}^{x_{1}}dx\int_{r_{+}}^{r_{max}}dr\frac{r^{2}}{(\ell +x r)^{2}}(r_{*}(r)-r_{*}(r_{b}))
\end{equation}
\begin{eqnarray}
  \text{Vol}(w_{t})&=&\text{v}_{0}+4 \pi \Delta \int_{r_{m_{1}}}^{r_{m_{2}}}dr (r(t+r_{*}(r_{b})-r_{*}(r))) \\
   \text{v}_{0}&=&8 \pi \Delta\int_{r_{+}}^{r_{max}}dr (r (r_{*}(r)-r_{*}(r_{b})))
   \end{eqnarray}
   
The contribution from Maxwell part of the action on the WDW patch
\begin{eqnarray}
   \text{Vol}(\mathcal{W}^{\text{F}}_{t})&=&V^{\text{F}}_{0}-16 q \ell^{2}\pi \Delta \int_{0}^{x_{1}}dx\int_{r_{m_{1}}}^{r_{m_{2}}}dr\frac{(\ell +x r)^{2}}{r^{2}}(t+r_{*}(r_{b})-r_{*}(r))\\
   V^{\text{F}}_{0}&=&-32 q \ell^{2}\pi \Delta \int_{0}^{x_{1}}dx\int_{r_{+}}^{r_{max}}dr\frac{(\ell +x r)^{2}}{r^{2}}(r_{*}(r)-r_{*}(r_{b}))
\end{eqnarray}

\noindent \textbf{Joint contribution}:\\
The intersection of the left and right past null boundaries with their corresponding outward pointing normal vector $n_{\mu}^{(L)} = \alpha_{1} \left[ dt - dr_* \right]_{\mu}$ and $n_{\mu}^{(R)} = - \alpha_{2} \left[ dt + dr_* \right]_{\mu}$ for $\alpha_1, \alpha_2 >0$. The inner product is 
\begin{align}
    n^{(L)} \cdot n^{(R)} =  {n^{(L)}}_{\mu} n^{(R)}{}^{\mu} = \frac{2 \alpha_1 \alpha_2 (\ell + r x)^2}{\ell^2 H(r)}
\end{align}
The joint action  is given by
\begin{align}
I_{\text{Joint}} = \frac{1}{8 \pi G_4} \int d^2 y \sqrt{\sigma} a \, \quad \text{with} \quad  a = \log \left(- n^{(L)}_{\mu} n^{(R)}{}^{\mu} / 2 \right).
\end{align}

We need to consider the contribution from joint 1 at $r_m^1$ and joint 2 at $r_m^2$ (Intersection of future and past null boundary in the black hole region \emph{i.e.} $r_- < r_m^i < r_+$  for $i \in (1,2)$ respectively). 

The contributions from joint 1 and joint 2 are given by,
\begin{eqnarray}
I_{\text{Joint}}^{J_{i}}=\frac{1}{8\pi G_{4}}\int_{0}^{x_{1}}dx\int_{0}^{2\pi \Delta}d\phi\left(\frac{\ell^{2}r^{2}}{(\ell+r x)^{2}}\log\left(\frac{\alpha_{1}\alpha_{2}(\ell+r x)}{\ell^{2}H(r)}\right)\right)_{r=r_{m_{i}}}
\end{eqnarray}
As the 3rd joint is at the boundary, it will have a diverging contribution, which needs to be regularized. This regularization has been done using a cutoff surface at $r=r_{max}$, and in this case the contribution  from $J_{3}$
\begin{eqnarray}
I_{\text{Joint}}^{J_{3}}=-\frac{1}{8\pi G_{4}}\int_{0}^{x_{1}}dx\int_{0}^{2\pi \Delta}d\phi\bigg(\frac{\ell^{2}r^{2}}{(\ell+r x)^{2}}\log\left(\frac{\alpha_{1}(\ell+r x)}{\ell\sqrt{H(r)}}\right) \\
    +\frac{\ell^{2}r^{2}}{(\ell+r x)^{2}}\log\left(\frac{\alpha_{1}(\ell+r x)}{\ell\sqrt{H(r)}}\right)\bigg)_{r=r_{max}}
\end{eqnarray}
The joint contributions from the surfaces $J_{4}$ and $J_{5}$ vanish by the same argument as for the regularized WDW patch.

From the expressions of all the contributing terms it is clear that their final contribution to the complexity gives rise to the same late time slope as we have got from the regularized WDW patch. Therefore it is justified to consider the regularized WDW patch $\mathcal{\tilde{W}}_{\tilde{t}}$ rather than full WDW patch $\mathcal{W}_{\tilde{t}}$.



\bibliographystyle{JHEP}
\bibliography{biblio.bib}

\end{document}